\documentclass[twocolumn,prl,aps,footinbib,amsmath,amssymb,superscriptaddress,longbibliography]{revtex4-1}
\usepackage{graphicx}
\usepackage{dcolumn}
\usepackage{xcolor}
\usepackage{bm}
\usepackage{natbib,hyperref}
\usepackage{amsmath,amssymb,amsfonts,bbold}
\usepackage[normalem]{ulem}
\usepackage{setspace}

\definecolor{dartmouthgreen}{rgb}{0.05, 0.5, 0.06}
\definecolor{darkspringgreen}{rgb}{0.09, 0.45, 0.27}
\definecolor{DebianRed}{rgb}{0.84, 0.04, 0.33}
\definecolor{darkpowderblue}{rgb}{0.0, 0.2, 0.6}

\newcommand{\beq}{\begin{equation}}
\newcommand{\eeq}{\end{equation}}
\newcommand{\beqa}{\begin{eqnarray}}
\newcommand{\eeqa}{\end{eqnarray}}

\begin{document}
\title{A Systematic Investigation of the Coupling between One-Dimensional Edge States \\ of a Topological Crystalline Insulator}
\author{Johannes Jung}
	\affiliation{Physikalisches Institut, Experimentelle Physik II, 
		Universit\"{a}t W\"{u}rzburg, Am Hubland, 97074 W\"{u}rzburg, Germany}
\author{Artem Odobesko}
	\email[corresponding author: \\]{artem.odobesko@physik.uni-wuerzburg.de}
	\affiliation{Physikalisches Institut, Experimentelle Physik II, 
		Universit\"{a}t W\"{u}rzburg, Am Hubland, 97074 W\"{u}rzburg, Germany}
\author{Robin Boshuis}
	\affiliation{Physikalisches Institut, Experimentelle Physik II, 
		Universit\"{a}t W\"{u}rzburg, Am Hubland, 97074 W\"{u}rzburg, Germany}
\author{Andrzej Szczerbakow} 
        \affiliation{Institute of Physics, Polish Academy of Sciences, 
        Aleja Lotnik\'ow 32/46, 02-668 Warsaw, Poland}
\author{Tomasz Story} 
        \affiliation{Institute of Physics, Polish Academy of Sciences, 
        Aleja Lotnik\'ow 32/46, 02-668 Warsaw, Poland}
        \affiliation{International Research Centre MagTop, Institute of Physics, 
        Polish Academy of Sciences, Aleja Lotnikow 32/46, 02-668 Warsaw, Poland}
\author{Matthias Bode} 
	\address{Physikalisches Institut, Experimentelle Physik II, 
	Universit\"{a}t W\"{u}rzburg, Am Hubland, 97074 W\"{u}rzburg, Germany}	
\date{\today}

\begin{abstract}
The interaction of spin-polarized one-dimensional (1D) topological edge modes 
localized along single-atomic steps of the topological crystalline insulator Pb$_{0.7}$Sn$_{0.3}$Se(001) 
has been studied systematically by scanning tunneling spectroscopy. 
Our results reveal that the coupling of adjacent edge modes sets in at a step--to--step distance $d_{\rm ss} \leq 25$\,nm, 
resulting in a characteristic splitting of a single peak at the Dirac point in tunneling spectra. 
Whereas the energy splitting exponentially increases with decreasing $d_{\rm ss}$ for single-atomic steps running almost parallel, 
we find no splitting for single-atomic step edges under an angle of $90^{\circ}$. 
The results are discussed in terms of overlapping wave functions with $p_x, p_y$ orbital character.
\end{abstract}

\maketitle

{\em Introduction ---}
The discovery of two-dimensional (2D) topological insulators (TIs), such as HgTe/CdTe\,\cite{Koenig2007} 
or InAs/GaSb quantum wells\,\cite{PhysRevLett.107.136603}, generated significant attention, 
especially due to the existence of edge states protected by time-reversal symmetry.  
The spin--momentum locking pertinent to these states mandates that charge carriers with a given spin 
can only propagate in a predetermined direction, thereby inhibiting disorder scattering 
and greatly reducing dissipation in electrical transport\,\cite{Bauer2016,Hou2019,Liu2019,Plucinski_2019,Liu_2020}.  
Further material classes with topologically protected band structures have been predicted and realized, 
such as topological crystalline insulators (TCI) \,\cite{Fu2011,Dziawa2012,Xu2012,Tanaka2012} and  
Weyl semimetals\,\cite{PhysRevX.5.011029,XBA2015,LWF2015,YLS2015,LXW2015,PhysRevB.95.035114}. 
However, although the ruggedness of the respective topological states is often praised as one of their main advantages, 
the reliable realization of one-dimensional (1D) topological conductance channels 
highly relevant for practical applications remains a challenge.  
Some systems with topological 1D states, like higher-order topological hinge states \cite{Drozdov2014,Schindler2018}, 
or surface step edges of weak TIs \cite{Pauly2015} or TCIs \cite{Sessi2016}, could be identified, but
systematic studies which address the interaction of adjacent channels remain elusive.  
As a result the question up to which density topological 1D channels may be packed 
without compromising the signal quality has not yet been answered.

In this context hybridization effects of topological states play a particularly important role. 
For example, the close proximity to the underlying substrate may lead to a significant interaction 
of topological edge states with substrate bulk states\,\cite{PhysRevLett.107.166801,PhysRevLett.109.016801,PhysRevB.89.155436}. 
Calculations predict that---if tuned appropriately---these effects may even be utilized, 
such as in topological-insulator--ferromagnetic-metal (TI--FM) heterostructures, 
where hybridization-induced interface states lead to a large spin-transfer torque\,\cite{PhysRevB.96.235433}.
Strong hybridization between the top and the bottom surfaces, however, 
can also result in band crossings resulting in a trivial TCI phase\,\cite{LHW2014}
or---for very thin films of three-dimensional TIs---to the complete quenching of the topological surface state\,\cite{Zhang2010}.  

In this contribution we systematically study the interactions of one-dimensional topological states 
which arise at odd-atomic step edges of (001)-terminated (Pb,Sn)Se single crystals.  
Whereas the rock salt structure compound PbSe is a trivial insulator, 
the order of p-derived cation and anion surface states bands is inverted 
for Pb$_{1-x}$Sn$_{x}$Se at $x \gtrsim 0.2$, i.e., when substituting 
a sufficient fraction of Pb atoms with Sn\,\cite{Dziawa2012,PhysRevB.90.161202}. 
Band inversion results in four Dirac cones per surface Brillouin zone (SBZ)
centered in close proximity to the $\overline{X}$ and $\overline{Y}$ points 
which are protected by $\langle 110\rangle$ mirror symmetry\,\cite{Hsieh2012,Liu2013,Wang2013,Safaei2013}.  

Clean (001) surfaces of Pb$_{1-x}$Sn$_{x}$Se are typically created by cleaving bulk crystals.
As first shown in Ref.\,\onlinecite{Okada2013}, tunneling spectra of flat (Pb,Sn)Se (001) surfaces 
exhibit a V-shaped dip with two symmetrically arranged side peaks which were assigned to the Dirac point and two 
Van Hove singularities caused by saddle points in the dispersion, respectively.  
However, the cleaving process unavoidably also results in a variety of steps of different height and shape.
Whereas the electronic structure of double- or other even-atomic step edges 
is virtually identical to a flat terrace, see Ref.\,\cite{SupplMat} (Note 1), 
the observation of a 1D topological edge mode at odd-atomic step edges, e.g., steps with a height 
of one or three atomic layers, which are characterized by a strong peak in the LDOS at the Dirac energy, 
was reported in Ref.\,\onlinecite{Sessi2016} and confirmed in Ref.\,\onlinecite{Madhavan2019}.
As detailed in Ref.\,\onlinecite{Rechcinski2018}, adjacent extended (001) terraces of (Pb,Sn)Se
must---due to the displacement by one atomic layer in a two-atomic unit cell---be described 
by Hamiltonians with chiral symmetry and an inverted order of states. 
The resulting surface states are divided into two classes with an even or odd number of layers.
Therefore, odd-atomic step edges host a doubly-degenerate flat band 
at the Dirac energy which is protected by the mirror plane symmetry of the film\,\cite{Rechcinski2018}. 

In this paper we systematically investigate the interaction 
between closely adjacent 1D edge states of the (001) surface of the TCI (Pb,Sn)Se. 
Hybridization is observed for step--to--step distances $d_{\rm ss} \lesssim 25$\,nm, 
resulting in the splitting of the single peak in tunneling spectra at the Dirac point into a double-peak. 
Our results reveal that the energetic splitting exponentially increases with decreasing $d_{\rm ss}$.
It is largely insensitive to the details of a particular configuration of step edges as long as the steps run approximately in parallel. 
In contrast, for single-atomic steps crossing at an almost perpendicular angle, we do not observe any splitting.  

{\em Methods ---}
Experiments have been performed in an ultra-high vacuum system 
equipped with a cryogenic STM (operation temperature $T = 4.9$\,K). 
Pb$_{0.7}$Sn$_{0.3}$Se p-doped single crystals grown by the self-selecting vapor growth 
method\,\cite{Dziawa2012,Sessi2016} (see Ref.\ \onlinecite{SupplMat}, Note 2, for further information) 
have been cleaved in UHV at a base pressure $p < 10^{-10}$\,mbar and immediately transferred into the STM. 
Tunneling conductance $\mathrm{d}I/\mathrm{d}U$ maps and spectra were measured by lock-in technique with 
modulation voltage $U_{\textrm{mod}} = 2.5$\,mV at a frequency $f = 790$\,Hz. 

{\em Results ---}
\begin{figure}[b]
    \includegraphics[width=\columnwidth]{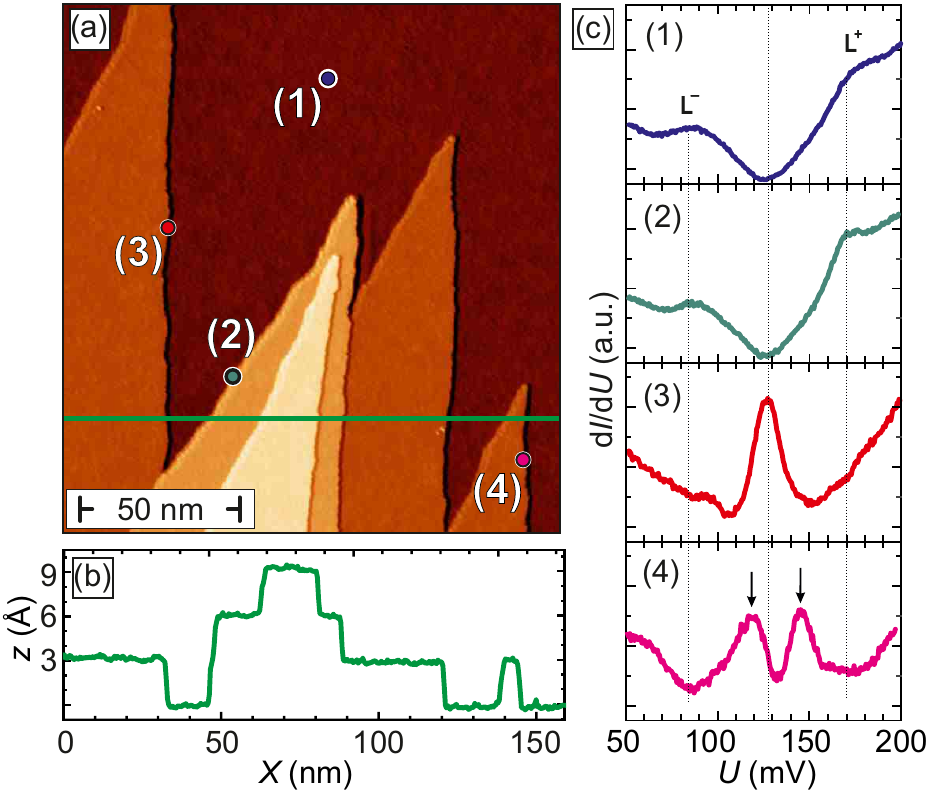}
	\caption{(a) Topographic STM image of a cleaved Pb$_{0.7}$Sn$_{0.3}$Se surface 
		with a number of wedge-shaped terraces with a single- and double-atomic step edge height;
		(b) A line profile measured along the green line in (a); 
		(c) Typical tunneling spectra measured on the plain terrace (1), double- (2) and single-atomic (3) step edges. 
		Spectrum (4) with a characteristic double-peak was obtained 
		at a single-atomic step edge close to the pointed end of of the wedge-shaped plateau. 
		Stabilization parameters: $U_{\textrm{set}} = 200$\,mV, $I_{\textrm{set}} = 0.2$\,nA.}
	\label{fig:S&D}             
\end{figure}
Figure \ref{fig:S&D}(a) shows a typical STM topography image of the cleaved (001) surface of the TCI  Pb$_{0.7}$Sn$_{0.3}$Se. 
Numerous wedge-shaped atomically flat pla\-teaus can be recognized. 
The line profile measured along the green line in Fig.\,\ref{fig:S&D}(a), presented in Fig.\,\ref{fig:S&D}(b),
reveals that most step edges are about 3\,{\AA} high, corresponding to half of the two-atomic unit cell.
Fig.\,\ref{fig:S&D}(c) shows typical $\mathrm{d}I/\mathrm{d}U$ spectra 
measured on a flat terrace far away from any step edge (1), 
on a double-atomic step edge (2), and at two single-atomic step edges [points (3),(4)]. 
Spectrum (1) exhibits a minimum at the Dirac point $E_{\rm D} = (125 \pm 5)$\,meV and two shoulders (L$^{-}$ and L$^{+}$), 
which are associated with the two saddle points in the dispersion\,\cite{Okada2013,Liu2013}. 
Spectrum (2) taken on a double-step edge is virtually indistinguishable from (1), 
indicating the absence of any edge state at step edges equivalent to an even number of atomic layers. 
In agreement with our earlier publication\,\cite{Sessi2016}, a pronounced peak at the Dirac energy 
indicative for the 1D edge state can only be found at odd-atomic step edges, 
such as spectrum (3) measured on a single-atomic step edge.
However, we find that the spectra of single-atomic step edges 
at some locations deviate from this simple single-peak structure. 
For example, the spectrum measured at (4) close to the termination point 
of an acute-angled wedge-shaped plateau reveals a pronounced double-peak feature 
in the vicinity of the Dirac point with a peak splitting of about $25$\,mV. 

\begin{figure*}[t]
	\includegraphics[width=1\textwidth]{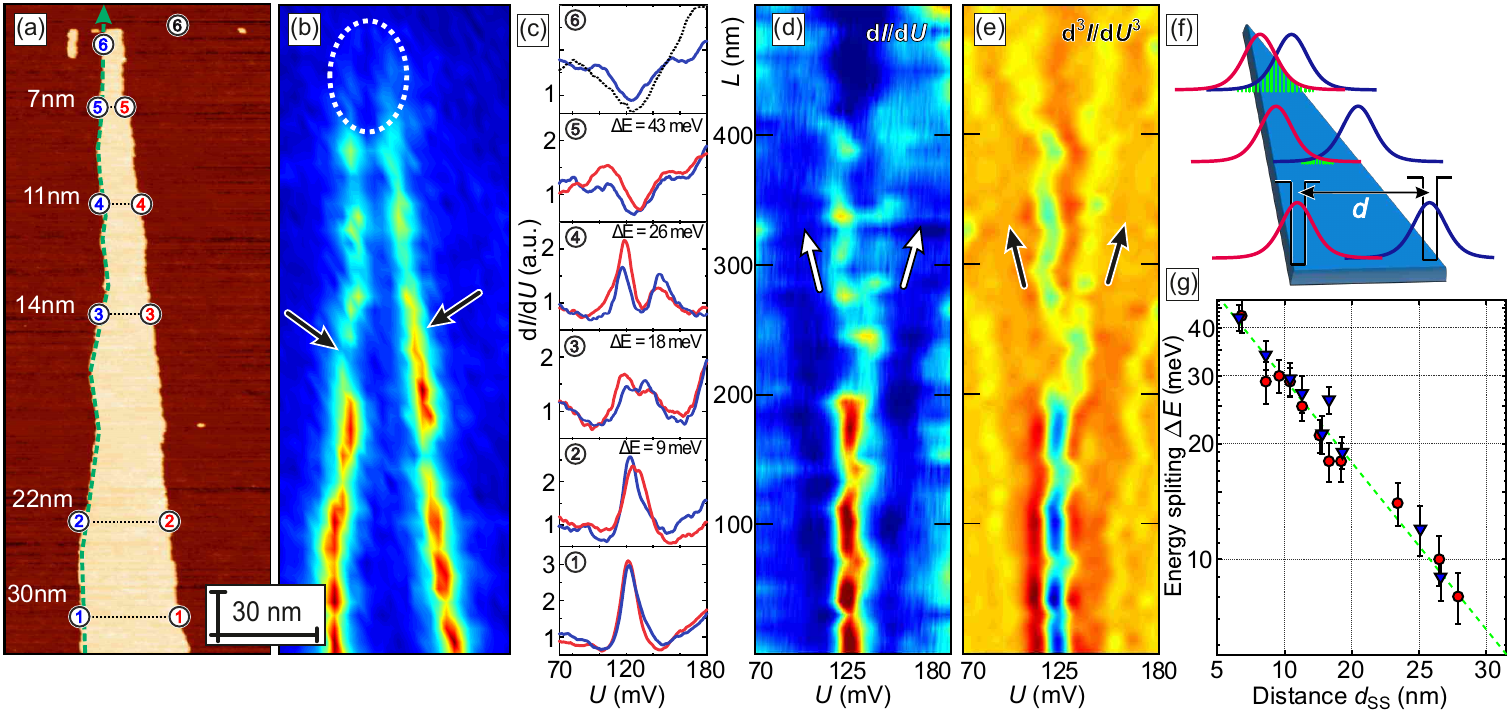}
	\caption{(a) Topographic STM image of a wedge-shaped plateau surrounded by single-atomic step edges.  
		Note, that the $x$- and $y$-direction are scaled differently, i.e., the wedge shape is even more acute-angled than it appears. 
		(b) $\mathrm{d}I/\mathrm{d}U$ map measured at the Dirac energy of the same region shown in (a).  
		An enhanced intensity at the position of the step edges is clearly visible in the bottom part, 
		indicating the presence of the topologically protected edge state. 
		As the wedge-shaped plateau becomes more narrow, the intensity of the $\mathrm{d}I/\mathrm{d}U$ signal 
		measured at both step edges decreases (see region above arrows) 
		and eventually vanishes close to the termination point (ellipse).
		(c) Tunneling spectra measured at the left (blue) and right (red) 
		single-atomic step edge at various step--to--step distances indicated in (a).		
		(d) Color-coded raw $\mathrm{d}I/\mathrm{d}U$ spectra measured along the green hatched line in (a).
		(e) Numerically calculated second derivative ($\mathrm{d^3}I/\mathrm{d}U^3$) of (d).  
		(f) Sketch of a model of two quantum wells localized along two converging single-atomic step edges. 
		The wave-functions extend beyond the walls and overlap. 
		The overlap area (green) indicates the coupling strength and determines the energy splitting.
		(g) The peak energy splitting $\Delta E$ in dependence of step--to--step distance im semilogarithmic scale. 
		Stabilization parameters: (a,b) $U_{\textrm{set}} = 125$\,mV (a), $I_{\textrm{set}} = 0.2$\,nA.}		
		\label{fig:DistDep}
\end{figure*}

To understand the physical origin of this peak splitting we performed detailed spectroscopic investigations 
at numerous single-atomic step edge structures on a variety of topological Pb$_{0.7}$Sn$_{0.3}$Se surfaces. 
Our results indicate that the peak splitting observed in Fig.\,\ref{fig:S&D}(c) on the single-atomic step edge  
is not caused by the presence of kinks, defects, or random variations of the local doping level\,\cite{Beidenkopf2011}.
Instead, the data reveal that the splitting occurs when two single-atomic step edges are in close proximity. 
A particularly illustrative case is displayed in Fig.\,\ref{fig:DistDep}.  
The topographic STM image of Fig.\,\ref{fig:DistDep}(a) shows a wedge-shaped plateau 
bound by two single-atomic step edges which converge under an acute angle of $3.4^{\circ} \pm 0.3^{\circ}$.  
As a result, the step--to--step distance $d_{\rm ss}$ slowly decreases over a total length of about 500\,nm
from $d_{\rm ss} > 30$\,nm in the bottom part to almost zero at the apex of the wedge.  
Both steps have rough edges which clearly deviate from high symmetry directions. 
Nevertheless, the $\mathrm{d}I/\mathrm{d}U$ map measured at a bias voltage corresponding
to the Dirac point presented in Fig.\,\ref{fig:DistDep}(b) shows several unique features.
First, in the bottom part of Fig.\,\ref{fig:DistDep}(b) where the two single-atomic step edges are far apart 
an intense $\mathrm{d}I/\mathrm{d}U$ signal can be recognized along the step edges, indicating the presence of the topological edge state.  
Second, as $d_{\rm ss}$ falls below about 15\,nm, indicated by two arrows in Fig.\,\ref{fig:DistDep}(b), 
the intensity of the $\mathrm{d}I/\mathrm{d}U$ signal noticeably decreases.  
Third, close to the top of the wedge where $d_{\rm ss} \lesssim 7$\,nm (marked by a hatched ellipse) 
it becomes essentially indistinguishable from the signal measured on the surrounding terraces.   

This modification of the LDOS at the Dirac energy correlates with a systematic alteration of the local STS data.  
Some exemplary spectra measured at various step--to--step distances 
ranging from $d_{\rm ss} = 30$\,nm down to $d_{\rm ss} = 7$\,nm are shown in Fig.\,\ref{fig:DistDep}(c).  
In each case we present the spectra measured at the left and the right single-atomic step edge in blue and red, respectively. 
The tunneling spectra measured at $d_{\rm ss} = 30$\,nm (curve {\textcircled{\footnotesize 1}}) 
exhibit a single intense peak at the Dirac energy $E_{\rm D} = 125$\,meV 
which is fully consistent with the spectra measured at isolated step edges\,\cite{Sessi2016}.
At $d_{\rm ss} = 22$\,nm (curve {\textcircled{\footnotesize 2}}) the peak height slightly decreases 
and a weak shoulder (blue curve) or subtle double humps (red curve) can be recognized. 
With $d_{\rm ss}$ reduced to 14\,nm (curve {\textcircled{\footnotesize 3}}) 
a clear splitting in two peaks becomes apparent, which amounts to $\Delta E \approx 18$\,mV.  
The splitting further increases with decreasing step--to--step distance reaching 
$\Delta E \approx 26$\,meV at $d_{\rm ss} = 11$\,nm (curve {\textcircled{\footnotesize 4}}).  
At $d_{\rm ss} = 7$\,nm the peak intensities have drastically dropped
(curve {\textcircled{\footnotesize 5}}), even though the splitting still appears to increase ($\Delta E \approx 43$\,meV).  
Eventually, at even smaller $d_{\rm ss}$ the peaks become almost undistinguishable 
from the spectra of the terrace (curve {\textcircled{\footnotesize 6}} and dashed line, respectively). 

Figure\,\ref{fig:DistDep}(d) represents the complete data set of color-coded raw $\mathrm{d}I/\mathrm{d}U$ spectra 
measured along the green hatched line in Fig.\,\ref{fig:DistDep}(a).  
This line section follows the contour of the left single-atomic step edge for about 500\,nm 
from the bottom of Fig.\,\ref{fig:DistDep}(a) ($L = 0$) all the way 
to its termination point at the upper part of the the wedge-shaped plateau. 
For better contrast the numerically calculated second derivative of the $\mathrm{d}I/\mathrm{d}U$ signal, 
i.e., $\mathrm{d^3}I/\mathrm{d}U^3$, is presented in Fig.\,\ref{fig:DistDep}(e). 
At 0\,nm\,$\le L \le 200$\,nm the data confirm the existence of a single peak 
at the Dirac energy $E_{\rm D}=eU \approx 125$\,meV.
Local fluctuations by about $\pm 5$\,mV are probably caused by statistical variations 
of the substitutional dopant concentration\,\cite{Beidenkopf2011,Storz2016}. 
As indicated by arrows in Fig.\,\ref{fig:DistDep}(d) and (e), for $L \ge 250$\,nm the single peak 
which is known to represent the topological edge state of (Pb,Sn)Se\,\cite{Sessi2016} splits into two diverging branches. 
The data presented in Fig.\,\ref{fig:DistDep}(e) reveal the peak splitting increases 
with decreasing step--to--step distance up to $L \leq 400$\,nm where $\Delta E \approx 28$\,meV at $d_{\rm ss} \approx 10$\,nm. 
Above this line both peaks experience significant peak-broadening 
and strongly decrease in intensity, possibly due to the onset of attenuation. 
As presented in detail in Ref.\,\onlinecite{SupplMat} (Note 3), the results obtained on the right step edge 
are fully consistent with the results presented for the left step edge in Fig.\,\ref{fig:DistDep}. 
Furthermore, these main findings are independent of the specific type and shape of the step edges. 
For example, results presented in Ref.\,\onlinecite{SupplMat} (Notes 4 and 5) 
confirm that the peak splitting also exists for wedge-shaped vacancy islands 
and nearby parallel steps, respectively.  

{\em Discussion ---}
Similar observations of a coupling-dependent energy splitting have been made 
for zero-dimensional quantum dots\,\cite{Nilson2018}
and two-dimensional layered van der Waals heterostructures\,\cite{Diaz2015}.  
In the latter case the opening of band-gaps was particularly prominent for out-of-plane--oriented MoS$_2$ orbitals 
which strongly overlap with graphene $\pi$-bands\,\cite{Diaz2015}.  
In fact, periodic boundary calculations presented in Fig.\,6 of Ref.~\onlinecite{Rechcinski2018} suggest 
that the topological 1D edge state of (Pb,Sn)Se remains largely unaffected for terrace width $d_{\rm ss} \ge 86$\,nm, 
in reasonable agreement with our experimental data, cf.\ Fig.\,\ref{fig:DistDep}.
Similar to the tight binding calculations presented in Ref.\,\onlinecite{Sessi2016}, 
a weak splitting may be visible for $d_{\rm ss} \ge 43$\,nm\,\cite{Rechcinski2018}.  
For narrower terraces, however, a collapse of the double Dirac cone into a single cone 
with a crossing at the $\overline{X}$ point was predicted\,\cite{Rechcinski2018}, 
in clear disagreement with our results which show the persistence 
of a split peak well below $d_{\rm ss} \le 10$\,nm, cf.\ Fig.\,\ref{fig:DistDep}.

To simulate the observed behavior, we assume that the formation of the edge state results 
in an increased local density of states, crammed in a narrow region around the step edge. 
This regions of accumulated charge are---in a first-order approach---modeled 
by narrow identical quantum wells with a widths $\sim 1$\,nm perpendicular to the step edge 
and much larger along the step edge, determined as the intra-row coherence length.
For a perfect step edge the topological edge mode would be completely delocalized along the step edge. 
The fact that the energy splitting observed at different points along the step edge 
depends on the step--to--step distance suggests that we don't deal with a single, 
infinitely extended quantum state, but with electron wave functions which are localized to a certain extent 
due to their finite coherence length, potentially induced by disorder. 
Two quantum wells are separated by a distance $d$ and affect the edge state, see Fig.\,\ref{fig:DistDep}(f). 
The well depth is assumed to $\sim 100$\,meV,  corresponding to the bulk band gap of the Pb$_{0.7}$Sn$_{0.3}$Se
\footnote{ 
	We would like to note that we cannot exclude that similar or even better results 
	can be obtained by more elaborate models, where the electrostatic interaction becomes less relevant 
	and is instead replaced by a behavior governed by spatially varying topological indices}.
As shown in detail in Ref.\,\onlinecite{SupplMat} (Note 6), 
both quantum wells host wave functions which extend beyond their respective boundaries 
and decay exponentially with $\exp(-\frac{x}{\xi})$ outside the well, with the decay length $\xi$. 

When the wave functions overlap at low $d$, see green area in Fig.\,\ref{fig:DistDep}(f), 
the exponentially decaying tails of the wave functions couple, resulting in an energy splitting $\Delta E$.   
Indeed, a plot of $\Delta E$ values extracted from Fig.\,\ref{fig:DistDep}(e) versus $d_{\rm ss}$
confirms the expected exponential dependence, see Fig.\,\ref{fig:DistDep}(g). 
The decay length of the 1D edge mode perpendicular to the step direction is determined to $\xi = (10 \pm 1)$\,nm.
By taking $\hbar \delta k \propto \hbar /2 \xi$ and $\delta E \approx 10$\,meV as the width of the single peak 
in spectrum {\textcircled{\footnotesize 1}} of Fig.\,\ref{fig:DistDep}(c),
we estimate the group velocity to $\upsilon_{\rm gr} \approx \frac{\delta E}{\hbar \delta k} \approx 4.8 \cdot 10^4$\,m/s,
in good agreement with earlier band dispersion measurements\,\cite{Wang2013,Zeljkovic2014}.

\begin{figure}
    \includegraphics[width=1\columnwidth]{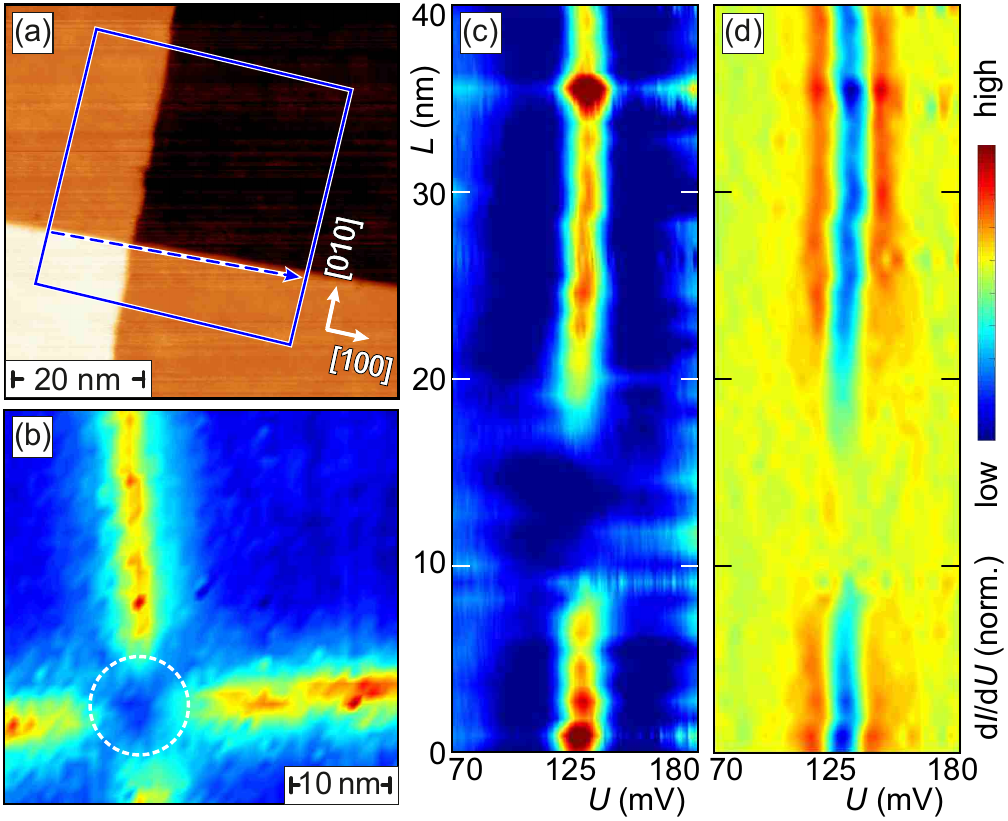}
	\caption{ (a) Topography and (b) $\mathrm{d}I/\mathrm{d}U$ map measured at the Dirac point ($U_{\rm set} = 120$ mV))
			of two intersecting at $90^{\circ}$ single-step edges, both running along the sample's high symmetry directions; 
			(c) Color-coded raw $\mathrm{d}I/\mathrm{d}U$ spectra measured along step-edge running in [100] marked 
			with the hatched line in (a); 
			(d) Second derivative ($\mathrm{d^3}I/\mathrm{d}U^3$) of the same signal in (c); 
			Stabilization parameters: $U_{\textrm{set}} = 120$\,mV, $I_{\textrm{set}} = 0.2$\,nA.}
	\label{fig:PerpStep}
\end{figure}
{\em Orthogonal step edges ---}
Our investigations of wedge-shaped islands like those presented in Figs.\,\ref{fig:S&D} and \ref{fig:DistDep} 
suggest that wedge angles are always $\lesssim 30^{\circ}$. 
In contrast, much larger adjacent angles we regularly observed at points where two single-atomic steps intersect.  
For example, the STM data of Fig.\,\ref{fig:PerpStep}(a) show two steps oriented along the planar 
high-symmetry $\langle 100 \rangle$ directions intersecting under an angle of $\approx 90^{\circ}$.  
Again, we find a strongly enhanced $\mathrm{d}I/\mathrm{d}U$ signal 
at the Dirac energy at the location of odd-atomic step edges and  
a strong reduction at the step crossing point, Fig.\,\ref{fig:PerpStep}(b).  
In contrast to the situation observed for wedge-shaped step edges in Fig.\,\ref{fig:DistDep}, however, 
the tunneling spectra measured along the white arrow in panel Fig.\,\ref{fig:PerpStep}(a) show no hints 
for any peak splitting, see Fig.\,\ref{fig:PerpStep}(c) and (d) for raw spectra and $\mathrm{d^3}I/\mathrm{d}U^3$, respectively.  
The full data set is presented in Ref.\,\,\onlinecite{SupplMat} (Note 7).

The results presented in Figs.\,\ref{fig:DistDep}, \ref{fig:PerpStep} and Ref.\,\onlinecite{SupplMat} 
show that the topological edge state characteristic for odd-atomic step edges of the (Pb,Sn)Se (001) surface
splits for adjacent steps ($d_{\rm ss} \lesssim 25$\,nm) running in parallel or under an acute angle.  
In contrast, for step edges crossing perpendicularly no such peak splitting is observed.  
This latter behavior hints at wave functions which are orthogonal such that the solution 
of the Schr{\"o}dinger equation becomes $\int \Phi^{\ast}_{n} \Phi_{m}dV = \delta_{n,m}$, 
where the indices $n,m$ stand for quantum numbers and $\delta_{n,m}$ is the Kronecker symbol.  
Obvious choices for this quantum number might be the spin or the orbital moment of the edge state.  
We are convinced that we can exclude the spin since results obtained 
within a tight binding\,\cite{Sessi2016} or envelope function model\,\cite{Rechcinski2018}
indicate an out-of-plane or vanishing spin polarization of the edge state, respectively, 
both insufficient to explain the absence of a peak splitting for orthogonal edge states. 
In contrast, the contribution of a Se $p_x$ and $p_y$ orbital momentum Dirac points 
at $\overline X$ and $\overline Y$ has been described theoretically\,%
\cite{PhysRevB.87.115106,Zeljkovic2014,Rechcinski2018} and nicely corresponds with our experimental findings.  
In order to test this interpretation we investigated intersections of two single-atomic steps at angles below but close to $90^{\circ}$.  
Indeed, for single-atomic steps that cross under a $60^{\circ}$ angle
we observe a very weak but detectable splitting, see Ref.\,\onlinecite{SupplMat} (Note 8). 
It is not clear, however, if the spin-orbit coupling in (Pb,Sn)Se plays any role.   
This question remains to be clarified by future theoretical investigations.  

{\em Conclusion---}
In conclusion, we have studied the interaction of adjacent 1D spin-polarized edge states 
which are localized at single-atomic step-edges of the TCI Pb$_{0.7}$Sn$_{0.3}$Se(001). 
Coupling at a step--to-step distance $d_{\rm ss} \lesssim 25$\,nm leads to the energy splitting 
of the single peak in LDOS at the Dirac point into two. 
The energy splitting increases exponentially with the reducing $d_{\rm ss}$ and eveals 
the localized character of the topological edge mode, instead of single quantum state.
At a very small separation ($d_{\rm ss}\lesssim10$\,nm) the two peaks strongly decrease in intensity.  
No peak splitting is found for perpendicular single-atomic step edges, 
suggesting an important role of the Se $p_x$ and $p_y$ orbital momentum of the edge state.

\begin{acknowledgments}
We acknowledge the contribution of P.\ Dziawa, J.\ Korczak, and R.\ Minikayev to crystal growth and characterization.
The work was supported by the DFG through SFB 1180 (project A02).  
We acknowledge financial support by the Deutsche Forschungsgemeinschaft (DFG, German Research Foundation) 
under Germany's Excellence Strategy through W{\"u}rzburg-Dresden Cluster of Excellence 
on Complexity and Topology in Quantum Matter -- ct.qmat (EXC 2147, project-id 390858490).
TS acknowledges the support from Foundation for Polish Science 
through IRA Programme co-financed by EU within Smart Growth Operational Programme.
\end{acknowledgments}


\begin{thebibliography}{43}%
\makeatletter
\providecommand \@ifxundefined [1]{%
 \@ifx{#1\undefined}
}%
\providecommand \@ifnum [1]{%
 \ifnum #1\expandafter \@firstoftwo
 \else \expandafter \@secondoftwo
 \fi
}%
\providecommand \@ifx [1]{%
 \ifx #1\expandafter \@firstoftwo
 \else \expandafter \@secondoftwo
 \fi
}%
\providecommand \natexlab [1]{#1}%
\providecommand \enquote  [1]{``#1''}%
\providecommand \bibnamefont  [1]{#1}%
\providecommand \bibfnamefont [1]{#1}%
\providecommand \citenamefont [1]{#1}%
\providecommand \href@noop [0]{\@secondoftwo}%
\providecommand \href [0]{\begingroup \@sanitize@url \@href}%
\providecommand \@href[1]{\@@startlink{#1}\@@href}%
\providecommand \@@href[1]{\endgroup#1\@@endlink}%
\providecommand \@sanitize@url [0]{\catcode `\\12\catcode `\$12\catcode
  `\&12\catcode `\#12\catcode `\^12\catcode `\_12\catcode `\%12\relax}%
\providecommand \@@startlink[1]{}%
\providecommand \@@endlink[0]{}%
\providecommand \url  [0]{\begingroup\@sanitize@url \@url }%
\providecommand \@url [1]{\endgroup\@href {#1}{\urlprefix }}%
\providecommand \urlprefix  [0]{URL }%
\providecommand \Eprint [0]{\href }%
\providecommand \doibase [0]{http://dx.doi.org/}%
\providecommand \selectlanguage [0]{\@gobble}%
\providecommand \bibinfo  [0]{\@secondoftwo}%
\providecommand \bibfield  [0]{\@secondoftwo}%
\providecommand \translation [1]{[#1]}%
\providecommand \BibitemOpen [0]{}%
\providecommand \bibitemStop [0]{}%
\providecommand \bibitemNoStop [0]{.\EOS\space}%
\providecommand \EOS [0]{\spacefactor3000\relax}%
\providecommand \BibitemShut  [1]{\csname bibitem#1\endcsname}%
\let\auto@bib@innerbib\@empty
\bibitem [{\citenamefont {K{\"o}nig}\ \emph {et~al.}(2007)\citenamefont
  {K{\"o}nig}, \citenamefont {Wiedmann}, \citenamefont {Br{\"u}ne},
  \citenamefont {Roth}, \citenamefont {Buhmann}, \citenamefont {Molenkamp},
  \citenamefont {Qi},\ and\ \citenamefont {Zhang}}]{Koenig2007}%
  \BibitemOpen
  \bibfield  {author} {\bibinfo {author} {\bibfnamefont {M.}~\bibnamefont
  {K{\"o}nig}}, \bibinfo {author} {\bibfnamefont {S.}~\bibnamefont {Wiedmann}},
  \bibinfo {author} {\bibfnamefont {C.}~\bibnamefont {Br{\"u}ne}}, \bibinfo
  {author} {\bibfnamefont {A.}~\bibnamefont {Roth}}, \bibinfo {author}
  {\bibfnamefont {H.}~\bibnamefont {Buhmann}}, \bibinfo {author} {\bibfnamefont
  {L.~W.}\ \bibnamefont {Molenkamp}}, \bibinfo {author} {\bibfnamefont {X.-L.}\
  \bibnamefont {Qi}}, \ and\ \bibinfo {author} {\bibfnamefont {S.-C.}\
  \bibnamefont {Zhang}},\ }\bibfield  {title} {\enquote {\bibinfo {title}
  {Quantum spin hall insulator state in {HgTe} quantum wells},}\ }\href
  {\doibase 10.1126/science.1148047} {\bibfield  {journal} {\bibinfo  {journal}
  {Science}\ }\textbf {\bibinfo {volume} {318}},\ \bibinfo {pages} {766--770}
  (\bibinfo {year} {2007})}\BibitemShut {NoStop}%
\bibitem [{\citenamefont {Knez}\ \emph {et~al.}(2011)\citenamefont {Knez},
  \citenamefont {Du},\ and\ \citenamefont {Sullivan}}]{PhysRevLett.107.136603}%
  \BibitemOpen
  \bibfield  {author} {\bibinfo {author} {\bibfnamefont {I.}~\bibnamefont
  {Knez}}, \bibinfo {author} {\bibfnamefont {R.-R.}\ \bibnamefont {Du}}, \ and\
  \bibinfo {author} {\bibfnamefont {G.}~\bibnamefont {Sullivan}},\ }\bibfield
  {title} {\enquote {\bibinfo {title} {Evidence for helical edge modes in
  inverted $\mathrm{InAs}/\mathrm{GaSb}$ quantum wells},}\ }\href {\doibase
  10.1103/PhysRevLett.107.136603} {\bibfield  {journal} {\bibinfo  {journal}
  {Phys. Rev. Lett.}\ }\textbf {\bibinfo {volume} {107}},\ \bibinfo {pages}
  {136603} (\bibinfo {year} {2011})}\BibitemShut {NoStop}%
\bibitem [{\citenamefont {Bauer}\ and\ \citenamefont
  {Bobisch}(2016)}]{Bauer2016}%
  \BibitemOpen
  \bibfield  {author} {\bibinfo {author} {\bibfnamefont {S.}~\bibnamefont
  {Bauer}}\ and\ \bibinfo {author} {\bibfnamefont {C.~A.}\ \bibnamefont
  {Bobisch}},\ }\bibfield  {title} {\enquote {\bibinfo {title} {Nanoscale
  electron transport at the surface of a topological insulator},}\ }\href
  {\doibase 10.1038/ncomms11381} {\bibfield  {journal} {\bibinfo  {journal}
  {Nature Comm.}\ }\textbf {\bibinfo {volume} {7}},\ \bibinfo {pages} {11381}
  (\bibinfo {year} {2016})}\BibitemShut {NoStop}%
\bibitem [{\citenamefont {Hou}\ \emph {et~al.}(2019)\citenamefont {Hou},
  \citenamefont {Wang}, \citenamefont {Xiao}, \citenamefont {McClintock},
  \citenamefont {Tra\-vaglini}, \citenamefont {Francia}, \citenamefont
  {Fetsch}, \citenamefont {Erten}, \citenamefont {Savrasov}, \citenamefont
  {Wang}, \citenamefont {Rossi}, \citenamefont {Vishik}, \citenamefont
  {Rotenberg},\ and\ \citenamefont {Yu}}]{Hou2019}%
  \BibitemOpen
  \bibfield  {author} {\bibinfo {author} {\bibfnamefont {Y.}~\bibnamefont
  {Hou}}, \bibinfo {author} {\bibfnamefont {R.}~\bibnamefont {Wang}}, \bibinfo
  {author} {\bibfnamefont {R.}~\bibnamefont {Xiao}}, \bibinfo {author}
  {\bibfnamefont {L.}~\bibnamefont {McClintock}}, \bibinfo {author}
  {\bibfnamefont {H.~C.}\ \bibnamefont {Tra\-vaglini}}, \bibinfo {author}
  {\bibfnamefont {J.~P.}\ \bibnamefont {Francia}}, \bibinfo {author}
  {\bibfnamefont {H.}~\bibnamefont {Fetsch}}, \bibinfo {author} {\bibfnamefont
  {O.}~\bibnamefont {Erten}}, \bibinfo {author} {\bibfnamefont {S.~Y.}\
  \bibnamefont {Savrasov}}, \bibinfo {author} {\bibfnamefont {B.}~\bibnamefont
  {Wang}}, \bibinfo {author} {\bibfnamefont {A.}~\bibnamefont {Rossi}},
  \bibinfo {author} {\bibfnamefont {I.}~\bibnamefont {Vishik}}, \bibinfo
  {author} {\bibfnamefont {E.}~\bibnamefont {Rotenberg}}, \ and\ \bibinfo
  {author} {\bibfnamefont {D.}~\bibnamefont {Yu}},\ }\bibfield  {title}
  {\enquote {\bibinfo {title} {Millimetre-long transport of photogenerated
  carriers in topological insulators},}\ }\href
  {https://doi.org/10.1038/s41467-019-13711-3} {\bibfield  {journal} {\bibinfo
  {journal} {Nature Comm.}\ }\textbf {\bibinfo {volume} {10}},\ \bibinfo
  {pages} {5723} (\bibinfo {year} {2019})}\BibitemShut {NoStop}%
\bibitem [{\citenamefont {Liu}\ \emph {et~al.}(2019)\citenamefont {Liu},
  \citenamefont {Williams},\ and\ \citenamefont {Cha}}]{Liu2019}%
  \BibitemOpen
  \bibfield  {author} {\bibinfo {author} {\bibfnamefont {P.}~\bibnamefont
  {Liu}}, \bibinfo {author} {\bibfnamefont {J.~R.}\ \bibnamefont {Williams}}, \
  and\ \bibinfo {author} {\bibfnamefont {J.~J.}\ \bibnamefont {Cha}},\
  }\bibfield  {title} {\enquote {\bibinfo {title} {Topological
  nanomaterials},}\ }\href {\doibase 10.1038/s41578-019-0113-4} {\bibfield
  {journal} {\bibinfo  {journal} {Nature Rev. Mater.}\ }\textbf {\bibinfo
  {volume} {4}},\ \bibinfo {pages} {479--496} (\bibinfo {year}
  {2019})}\BibitemShut {NoStop}%
\bibitem [{\citenamefont {Plucinski}(2019)}]{Plucinski_2019}%
  \BibitemOpen
  \bibfield  {author} {\bibinfo {author} {\bibfnamefont {L.}~\bibnamefont
  {Plucinski}},\ }\bibfield  {title} {\enquote {\bibinfo {title} {Band
  structure engineering in {3D} topological insulators},}\ }\href {\doibase
  10.1088/1361-648x/ab052c} {\bibfield  {journal} {\bibinfo  {journal} {Jour.
  Phys.: Cond. Matter}\ }\textbf {\bibinfo {volume} {31}},\ \bibinfo {pages}
  {183001} (\bibinfo {year} {2019})}\BibitemShut {NoStop}%
\bibitem [{\citenamefont {Liu}\ \emph {et~al.}(2020)\citenamefont {Liu},
  \citenamefont {Wang}, \citenamefont {Qiu},\ and\ \citenamefont
  {Gao}}]{Liu_2020}%
  \BibitemOpen
  \bibfield  {author} {\bibinfo {author} {\bibfnamefont {C.-W.}\ \bibnamefont
  {Liu}}, \bibinfo {author} {\bibfnamefont {Z.}~\bibnamefont {Wang}}, \bibinfo
  {author} {\bibfnamefont {R.~L.~J.}\ \bibnamefont {Qiu}}, \ and\ \bibinfo
  {author} {\bibfnamefont {X.~P.~A.}\ \bibnamefont {Gao}},\ }\bibfield  {title}
  {\enquote {\bibinfo {title} {Development of topological insulator and
  topological crystalline insulator nanostructures},}\ }\href {\doibase
  10.1088/1361-6528/ab6dfc} {\bibfield  {journal} {\bibinfo  {journal}
  {Nanotechn.}\ }\textbf {\bibinfo {volume} {31}},\ \bibinfo {pages} {192001}
  (\bibinfo {year} {2020})}\BibitemShut {NoStop}%
\bibitem [{\citenamefont {Fu}(2011)}]{Fu2011}%
  \BibitemOpen
  \bibfield  {author} {\bibinfo {author} {\bibfnamefont {L.}~\bibnamefont
  {Fu}},\ }\bibfield  {title} {\enquote {\bibinfo {title} {Topological
  crystalline insulators},}\ }\href {\doibase 10.1103/PhysRevLett.106.106802}
  {\bibfield  {journal} {\bibinfo  {journal} {Phys. Rev. Lett.}\ }\textbf
  {\bibinfo {volume} {106}},\ \bibinfo {pages} {106802} (\bibinfo {year}
  {2011})}\BibitemShut {NoStop}%
\bibitem [{\citenamefont {Dziawa}\ \emph {et~al.}(2012)\citenamefont {Dziawa},
  \citenamefont {Kowalski}, \citenamefont {Dybko}, \citenamefont {Buczko},
  \citenamefont {Szczerbakow}, \citenamefont {Szot}, \citenamefont
  {Lusakowska}, \citenamefont {Balasubramanian}, \citenamefont {Wojek},
  \citenamefont {Berntsen}, \citenamefont {Tjernberg},\ and\ \citenamefont
  {Story}}]{Dziawa2012}%
  \BibitemOpen
  \bibfield  {author} {\bibinfo {author} {\bibfnamefont {P.}~\bibnamefont
  {Dziawa}}, \bibinfo {author} {\bibfnamefont {B.~J.}\ \bibnamefont
  {Kowalski}}, \bibinfo {author} {\bibfnamefont {K.}~\bibnamefont {Dybko}},
  \bibinfo {author} {\bibfnamefont {R.}~\bibnamefont {Buczko}}, \bibinfo
  {author} {\bibfnamefont {A.}~\bibnamefont {Szczerbakow}}, \bibinfo {author}
  {\bibfnamefont {M.}~\bibnamefont {Szot}}, \bibinfo {author} {\bibfnamefont
  {E.}~\bibnamefont {Lusakowska}}, \bibinfo {author} {\bibfnamefont
  {T.}~\bibnamefont {Balasubramanian}}, \bibinfo {author} {\bibfnamefont
  {B.~M.}\ \bibnamefont {Wojek}}, \bibinfo {author} {\bibfnamefont {M.~H.}\
  \bibnamefont {Berntsen}}, \bibinfo {author} {\bibfnamefont {O.}~\bibnamefont
  {Tjernberg}}, \ and\ \bibinfo {author} {\bibfnamefont {T.}~\bibnamefont
  {Story}},\ }\bibfield  {title} {\enquote {\bibinfo {title} {Topological
  crystalline insulator states in {Pb$_{1-x}$Sn$_x$Se}},}\ }\href
  {https://doi.org/10.1038/nmat3449} {\bibfield  {journal} {\bibinfo  {journal}
  {Nature Mater.}\ }\textbf {\bibinfo {volume} {11}},\ \bibinfo {pages} {1023}
  (\bibinfo {year} {2012})}\BibitemShut {NoStop}%
\bibitem [{\citenamefont {Xu}\ \emph {et~al.}(2012)\citenamefont {Xu},
  \citenamefont {Liu}, \citenamefont {A.}, \citenamefont {Neupane},
  \citenamefont {Qian}, \citenamefont {Belopolski}, \citenamefont {Denlinger},
  \citenamefont {Wang}, \citenamefont {Lin}, \citenamefont {Wray},
  \citenamefont {Landolt}, \citenamefont {Slomski}, \citenamefont {Dil},
  \citenamefont {Marcinkova}, \citenamefont {Morosan}, \citenamefont {Gibson},
  \citenamefont {Sankar}, \citenamefont {Chou}, \citenamefont {Cava},
  \citenamefont {Bansil},\ and\ \citenamefont {Hasan}}]{Xu2012}%
  \BibitemOpen
  \bibfield  {author} {\bibinfo {author} {\bibfnamefont {S.-Y.}\ \bibnamefont
  {Xu}}, \bibinfo {author} {\bibfnamefont {C.}~\bibnamefont {Liu}}, \bibinfo
  {author} {\bibfnamefont {N.}~\bibnamefont {A.}}, \bibinfo {author}
  {\bibfnamefont {M.}~\bibnamefont {Neupane}}, \bibinfo {author} {\bibfnamefont
  {D.}~\bibnamefont {Qian}}, \bibinfo {author} {\bibfnamefont {I.}~\bibnamefont
  {Belopolski}}, \bibinfo {author} {\bibfnamefont {J.D.}\ \bibnamefont
  {Denlinger}}, \bibinfo {author} {\bibfnamefont {Y.J.}\ \bibnamefont {Wang}},
  \bibinfo {author} {\bibfnamefont {H.}~\bibnamefont {Lin}}, \bibinfo {author}
  {\bibfnamefont {L.A.}\ \bibnamefont {Wray}}, \bibinfo {author} {\bibfnamefont
  {G.}~\bibnamefont {Landolt}}, \bibinfo {author} {\bibfnamefont
  {B.}~\bibnamefont {Slomski}}, \bibinfo {author} {\bibfnamefont {J.H.}\
  \bibnamefont {Dil}}, \bibinfo {author} {\bibfnamefont {A.}~\bibnamefont
  {Marcinkova}}, \bibinfo {author} {\bibfnamefont {E.}~\bibnamefont {Morosan}},
  \bibinfo {author} {\bibfnamefont {Q.}~\bibnamefont {Gibson}}, \bibinfo
  {author} {\bibfnamefont {R.}~\bibnamefont {Sankar}}, \bibinfo {author}
  {\bibfnamefont {F.C.}\ \bibnamefont {Chou}}, \bibinfo {author} {\bibfnamefont
  {R.J.}\ \bibnamefont {Cava}}, \bibinfo {author} {\bibfnamefont
  {A.}~\bibnamefont {Bansil}}, \ and\ \bibinfo {author} {\bibfnamefont {M.Z.}\
  \bibnamefont {Hasan}},\ }\bibfield  {title} {\enquote {\bibinfo {title}
  {Observation of a topological crystalline insulator phase and topological
  phase transition in {{Pb}$_{1-x}${Sn}$_{x}${Te}}},}\ }\href
  {https://www.nature.com/articles/ncomms2191} {\bibfield  {journal} {\bibinfo
  {journal} {Nature Comm.}\ }\textbf {\bibinfo {volume} {3}},\ \bibinfo {pages}
  {1192} (\bibinfo {year} {2012})}\BibitemShut {NoStop}%
\bibitem [{\citenamefont {Tanaka}\ \emph {et~al.}(2012)\citenamefont {Tanaka},
  \citenamefont {Ren}, \citenamefont {Sato}, \citenamefont {Nakayama},
  \citenamefont {Souma}, \citenamefont {Takahashi}, \citenamefont {Segawa},\
  and\ \citenamefont {Ando}}]{Tanaka2012}%
  \BibitemOpen
  \bibfield  {author} {\bibinfo {author} {\bibfnamefont {Y.}~\bibnamefont
  {Tanaka}}, \bibinfo {author} {\bibfnamefont {Z.}~\bibnamefont {Ren}},
  \bibinfo {author} {\bibfnamefont {T.}~\bibnamefont {Sato}}, \bibinfo {author}
  {\bibfnamefont {K.}~\bibnamefont {Nakayama}}, \bibinfo {author}
  {\bibfnamefont {S.}~\bibnamefont {Souma}}, \bibinfo {author} {\bibfnamefont
  {T.}~\bibnamefont {Takahashi}}, \bibinfo {author} {\bibfnamefont
  {K.}~\bibnamefont {Segawa}}, \ and\ \bibinfo {author} {\bibfnamefont
  {Y.}~\bibnamefont {Ando}},\ }\bibfield  {title} {\enquote {\bibinfo {title}
  {Experimental realization of a topological crystalline insulator in
  {SnTe}},}\ }\href {https://www.nature.com/articles/nphys2442} {\bibfield
  {journal} {\bibinfo  {journal} {Nature Phys.}\ }\textbf {\bibinfo {volume}
  {8}},\ \bibinfo {pages} {800--803} (\bibinfo {year} {2012})}\BibitemShut
  {NoStop}%
\bibitem [{\citenamefont {Weng}\ \emph {et~al.}(2015)\citenamefont {Weng},
  \citenamefont {Fang}, \citenamefont {Fang}, \citenamefont {Bernevig},\ and\
  \citenamefont {Dai}}]{PhysRevX.5.011029}%
  \BibitemOpen
  \bibfield  {author} {\bibinfo {author} {\bibfnamefont {H.}~\bibnamefont
  {Weng}}, \bibinfo {author} {\bibfnamefont {C.}~\bibnamefont {Fang}}, \bibinfo
  {author} {\bibfnamefont {Z.}~\bibnamefont {Fang}}, \bibinfo {author}
  {\bibfnamefont {B.~A.}\ \bibnamefont {Bernevig}}, \ and\ \bibinfo {author}
  {\bibfnamefont {X.}~\bibnamefont {Dai}},\ }\bibfield  {title} {\enquote
  {\bibinfo {title} {{Weyl} semimetal phase in noncentrosymmetric
  transition-metal monophosphides},}\ }\href {\doibase
  10.1103/PhysRevX.5.011029} {\bibfield  {journal} {\bibinfo  {journal} {Phys.
  Rev. X}\ }\textbf {\bibinfo {volume} {5}},\ \bibinfo {pages} {011029}
  (\bibinfo {year} {2015})}\BibitemShut {NoStop}%
\bibitem [{\citenamefont {Xu}\ \emph {et~al.}(2015)\citenamefont {Xu},
  \citenamefont {Belopolski}, \citenamefont {Alidoust}, \citenamefont
  {Neupane}, \citenamefont {Bian}, \citenamefont {Zhang}, \citenamefont
  {Sankar}, \citenamefont {Chang}, \citenamefont {Yuan}, \citenamefont {Lee},
  \citenamefont {Huang}, \citenamefont {Zheng}, \citenamefont {Ma},
  \citenamefont {Sanchez}, \citenamefont {Wang}, \citenamefont {Bansil},
  \citenamefont {Chou}, \citenamefont {Shibayev}, \citenamefont {Lin},
  \citenamefont {Jia},\ and\ \citenamefont {Hasan}}]{XBA2015}%
  \BibitemOpen
  \bibfield  {author} {\bibinfo {author} {\bibfnamefont {S.-Y.}\ \bibnamefont
  {Xu}}, \bibinfo {author} {\bibfnamefont {I.}~\bibnamefont {Belopolski}},
  \bibinfo {author} {\bibfnamefont {N.}~\bibnamefont {Alidoust}}, \bibinfo
  {author} {\bibfnamefont {M.}~\bibnamefont {Neupane}}, \bibinfo {author}
  {\bibfnamefont {G.}~\bibnamefont {Bian}}, \bibinfo {author} {\bibfnamefont
  {C.}~\bibnamefont {Zhang}}, \bibinfo {author} {\bibfnamefont
  {R.}~\bibnamefont {Sankar}}, \bibinfo {author} {\bibfnamefont
  {G.}~\bibnamefont {Chang}}, \bibinfo {author} {\bibfnamefont
  {Z.}~\bibnamefont {Yuan}}, \bibinfo {author} {\bibfnamefont {C.-C.}\
  \bibnamefont {Lee}}, \bibinfo {author} {\bibfnamefont {S.-M.}\ \bibnamefont
  {Huang}}, \bibinfo {author} {\bibfnamefont {H.}~\bibnamefont {Zheng}},
  \bibinfo {author} {\bibfnamefont {J.}~\bibnamefont {Ma}}, \bibinfo {author}
  {\bibfnamefont {D.~S.}\ \bibnamefont {Sanchez}}, \bibinfo {author}
  {\bibfnamefont {B.~K.}\ \bibnamefont {Wang}}, \bibinfo {author}
  {\bibfnamefont {A.}~\bibnamefont {Bansil}}, \bibinfo {author} {\bibfnamefont
  {F.}~\bibnamefont {Chou}}, \bibinfo {author} {\bibfnamefont {P.~P.}\
  \bibnamefont {Shibayev}}, \bibinfo {author} {\bibfnamefont {H.}~\bibnamefont
  {Lin}}, \bibinfo {author} {\bibfnamefont {S.}~\bibnamefont {Jia}}, \ and\
  \bibinfo {author} {\bibfnamefont {M.~Z.}\ \bibnamefont {Hasan}},\ }\bibfield
  {title} {\enquote {\bibinfo {title} {Discovery of a {Weyl} fermion semimetal
  and topological {Fermi} arcs},}\ }\href {\doibase 10.1126/science.aaa9297}
  {\bibfield  {journal} {\bibinfo  {journal} {Science}\ }\textbf {\bibinfo
  {volume} {349}},\ \bibinfo {pages} {613--617} (\bibinfo {year}
  {2015})}\BibitemShut {NoStop}%
\bibitem [{\citenamefont {Lv}\ \emph {et~al.}(2015{\natexlab{a}})\citenamefont
  {Lv}, \citenamefont {Weng}, \citenamefont {Fu}, \citenamefont {Wang},
  \citenamefont {Miao}, \citenamefont {Ma}, \citenamefont {Richard},
  \citenamefont {Huang}, \citenamefont {Zhao}, \citenamefont {Chen},
  \citenamefont {Fang}, \citenamefont {Dai}, \citenamefont {Qian},\ and\
  \citenamefont {Ding}}]{LWF2015}%
  \BibitemOpen
  \bibfield  {author} {\bibinfo {author} {\bibfnamefont {B.~Q.}\ \bibnamefont
  {Lv}}, \bibinfo {author} {\bibfnamefont {H.~M.}\ \bibnamefont {Weng}},
  \bibinfo {author} {\bibfnamefont {B.~B.}\ \bibnamefont {Fu}}, \bibinfo
  {author} {\bibfnamefont {X.~P.}\ \bibnamefont {Wang}}, \bibinfo {author}
  {\bibfnamefont {H.}~\bibnamefont {Miao}}, \bibinfo {author} {\bibfnamefont
  {J.}~\bibnamefont {Ma}}, \bibinfo {author} {\bibfnamefont {P.}~\bibnamefont
  {Richard}}, \bibinfo {author} {\bibfnamefont {X.~C.}\ \bibnamefont {Huang}},
  \bibinfo {author} {\bibfnamefont {L.~X.}\ \bibnamefont {Zhao}}, \bibinfo
  {author} {\bibfnamefont {G.~F.}\ \bibnamefont {Chen}}, \bibinfo {author}
  {\bibfnamefont {Z.}~\bibnamefont {Fang}}, \bibinfo {author} {\bibfnamefont
  {X.}~\bibnamefont {Dai}}, \bibinfo {author} {\bibfnamefont {T.}~\bibnamefont
  {Qian}}, \ and\ \bibinfo {author} {\bibfnamefont {H.}~\bibnamefont {Ding}},\
  }\bibfield  {title} {\enquote {\bibinfo {title} {Experimental discovery of
  {Weyl} semimetal {TaAs}},}\ }\href {\doibase 10.1103/PhysRevX.5.031013}
  {\bibfield  {journal} {\bibinfo  {journal} {Phys. Rev. X}\ }\textbf {\bibinfo
  {volume} {5}},\ \bibinfo {pages} {031013} (\bibinfo {year}
  {2015}{\natexlab{a}})}\BibitemShut {NoStop}%
\bibitem [{\citenamefont {Yang}\ \emph {et~al.}(2015)\citenamefont {Yang},
  \citenamefont {Liu}, \citenamefont {Sun}, \citenamefont {Peng}, \citenamefont
  {Yang}, \citenamefont {Zhang}, \citenamefont {Zhou}, \citenamefont {Zhang},
  \citenamefont {Guo}, \citenamefont {Rahn}, \citenamefont {Prabhakaran},
  \citenamefont {Hussain}, \citenamefont {Mo}, \citenamefont {Felser},
  \citenamefont {Yan},\ and\ \citenamefont {Chen}}]{YLS2015}%
  \BibitemOpen
  \bibfield  {author} {\bibinfo {author} {\bibfnamefont {L.~X.}\ \bibnamefont
  {Yang}}, \bibinfo {author} {\bibfnamefont {Z.~K.}\ \bibnamefont {Liu}},
  \bibinfo {author} {\bibfnamefont {Y.}~\bibnamefont {Sun}}, \bibinfo {author}
  {\bibfnamefont {H.}~\bibnamefont {Peng}}, \bibinfo {author} {\bibfnamefont
  {H.~F.}\ \bibnamefont {Yang}}, \bibinfo {author} {\bibfnamefont
  {T.}~\bibnamefont {Zhang}}, \bibinfo {author} {\bibfnamefont
  {B.}~\bibnamefont {Zhou}}, \bibinfo {author} {\bibfnamefont {Y.}~\bibnamefont
  {Zhang}}, \bibinfo {author} {\bibfnamefont {Y.~F.}\ \bibnamefont {Guo}},
  \bibinfo {author} {\bibfnamefont {M.}~\bibnamefont {Rahn}}, \bibinfo {author}
  {\bibfnamefont {D.}~\bibnamefont {Prabhakaran}}, \bibinfo {author}
  {\bibfnamefont {Z.}~\bibnamefont {Hussain}}, \bibinfo {author} {\bibfnamefont
  {S.~K.}\ \bibnamefont {Mo}}, \bibinfo {author} {\bibfnamefont
  {C.}~\bibnamefont {Felser}}, \bibinfo {author} {\bibfnamefont
  {B.}~\bibnamefont {Yan}}, \ and\ \bibinfo {author} {\bibfnamefont {Y.~L.}\
  \bibnamefont {Chen}},\ }\bibfield  {title} {\enquote {\bibinfo {title}
  {{Weyl} semimetal phase in the non-centrosymmetric compound {TaAs}},}\ }\href
  {http://dx.doi.org/10.1038/nphys3425} {\bibfield  {journal} {\bibinfo
  {journal} {Nature Phys.}\ }\textbf {\bibinfo {volume} {11}},\ \bibinfo
  {pages} {728--732} (\bibinfo {year} {2015})}\BibitemShut {NoStop}%
\bibitem [{\citenamefont {Lv}\ \emph {et~al.}(2015{\natexlab{b}})\citenamefont
  {Lv}, \citenamefont {Xu}, \citenamefont {Weng}, \citenamefont {Ma},
  \citenamefont {Richard}, \citenamefont {Huang}, \citenamefont {Zhao},
  \citenamefont {Chen}, \citenamefont {Matt}, \citenamefont {Bisti},
  \citenamefont {Strocov}, \citenamefont {Mesot}, \citenamefont {Fang},
  \citenamefont {Dai}, \citenamefont {Qian}, \citenamefont {Shi},\ and\
  \citenamefont {Ding}}]{LXW2015}%
  \BibitemOpen
  \bibfield  {author} {\bibinfo {author} {\bibfnamefont {B.~Q.}\ \bibnamefont
  {Lv}}, \bibinfo {author} {\bibfnamefont {N.}~\bibnamefont {Xu}}, \bibinfo
  {author} {\bibfnamefont {H.~M.}\ \bibnamefont {Weng}}, \bibinfo {author}
  {\bibfnamefont {J.~Z.}\ \bibnamefont {Ma}}, \bibinfo {author} {\bibfnamefont
  {P.}~\bibnamefont {Richard}}, \bibinfo {author} {\bibfnamefont {X.~C.}\
  \bibnamefont {Huang}}, \bibinfo {author} {\bibfnamefont {L.~X.}\ \bibnamefont
  {Zhao}}, \bibinfo {author} {\bibfnamefont {G.~F.}\ \bibnamefont {Chen}},
  \bibinfo {author} {\bibfnamefont {C.~E.}\ \bibnamefont {Matt}}, \bibinfo
  {author} {\bibfnamefont {F.}~\bibnamefont {Bisti}}, \bibinfo {author}
  {\bibfnamefont {V.~N.}\ \bibnamefont {Strocov}}, \bibinfo {author}
  {\bibfnamefont {J.}~\bibnamefont {Mesot}}, \bibinfo {author} {\bibfnamefont
  {Z.}~\bibnamefont {Fang}}, \bibinfo {author} {\bibfnamefont {X.}~\bibnamefont
  {Dai}}, \bibinfo {author} {\bibfnamefont {T.}~\bibnamefont {Qian}}, \bibinfo
  {author} {\bibfnamefont {M.}~\bibnamefont {Shi}}, \ and\ \bibinfo {author}
  {\bibfnamefont {H.}~\bibnamefont {Ding}},\ }\bibfield  {title} {\enquote
  {\bibinfo {title} {Observation of {Weyl} nodes in {TaAs}},}\ }\href
  {http://dx.doi.org/10.1038/nphys3426} {\bibfield  {journal} {\bibinfo
  {journal} {Nature Phys.}\ }\textbf {\bibinfo {volume} {11}},\ \bibinfo
  {pages} {724--727} (\bibinfo {year} {2015}{\natexlab{b}})}\BibitemShut
  {NoStop}%
\bibitem [{\citenamefont {Sessi}\ \emph {et~al.}(2017)\citenamefont {Sessi},
  \citenamefont {Sun}, \citenamefont {Bathon}, \citenamefont {Glott},
  \citenamefont {Li}, \citenamefont {Chen}, \citenamefont {Guo}, \citenamefont
  {Chen}, \citenamefont {Schmidt}, \citenamefont {Felser}, \citenamefont
  {Yan},\ and\ \citenamefont {Bode}}]{PhysRevB.95.035114}%
  \BibitemOpen
  \bibfield  {author} {\bibinfo {author} {\bibfnamefont {P.}~\bibnamefont
  {Sessi}}, \bibinfo {author} {\bibfnamefont {Y.}~\bibnamefont {Sun}}, \bibinfo
  {author} {\bibfnamefont {T.}~\bibnamefont {Bathon}}, \bibinfo {author}
  {\bibfnamefont {F.}~\bibnamefont {Glott}}, \bibinfo {author} {\bibfnamefont
  {Z.}~\bibnamefont {Li}}, \bibinfo {author} {\bibfnamefont {H.}~\bibnamefont
  {Chen}}, \bibinfo {author} {\bibfnamefont {L.}~\bibnamefont {Guo}}, \bibinfo
  {author} {\bibfnamefont {X.}~\bibnamefont {Chen}}, \bibinfo {author}
  {\bibfnamefont {M.}~\bibnamefont {Schmidt}}, \bibinfo {author} {\bibfnamefont
  {C.}~\bibnamefont {Felser}}, \bibinfo {author} {\bibfnamefont
  {B.}~\bibnamefont {Yan}}, \ and\ \bibinfo {author} {\bibfnamefont
  {M.}~\bibnamefont {Bode}},\ }\bibfield  {title} {\enquote {\bibinfo {title}
  {Impurity screening and stability of {F}ermi arcs against {C}oulomb and
  magnetic scattering in a {W}eyl monopnictide},}\ }\href {\doibase
  10.1103/PhysRevB.95.035114} {\bibfield  {journal} {\bibinfo  {journal} {Phys.
  Rev. B}\ }\textbf {\bibinfo {volume} {95}},\ \bibinfo {pages} {035114}
  (\bibinfo {year} {2017})}\BibitemShut {NoStop}%
\bibitem [{\citenamefont {Drozdov}\ \emph {et~al.}(2014)\citenamefont
  {Drozdov}, \citenamefont {Alexandradinata}, \citenamefont {Jeon},
  \citenamefont {Nadj-Perge}, \citenamefont {Ji}, \citenamefont {Cava},
  \citenamefont {Bernevig},\ and\ \citenamefont {Yazdani}}]{Drozdov2014}%
  \BibitemOpen
  \bibfield  {author} {\bibinfo {author} {\bibfnamefont {I.~K.}\ \bibnamefont
  {Drozdov}}, \bibinfo {author} {\bibfnamefont {A.}~\bibnamefont
  {Alexandradinata}}, \bibinfo {author} {\bibfnamefont {S.}~\bibnamefont
  {Jeon}}, \bibinfo {author} {\bibfnamefont {S.}~\bibnamefont {Nadj-Perge}},
  \bibinfo {author} {\bibfnamefont {H.}~\bibnamefont {Ji}}, \bibinfo {author}
  {\bibfnamefont {R.~J.}\ \bibnamefont {Cava}}, \bibinfo {author}
  {\bibfnamefont {A.~B.}\ \bibnamefont {Bernevig}}, \ and\ \bibinfo {author}
  {\bibfnamefont {A.}~\bibnamefont {Yazdani}},\ }\bibfield  {title} {\enquote
  {\bibinfo {title} {One-dimensional topological edge states of bismuth
  bilayers},}\ }\href {\doibase 10.1038/nphys3048} {\bibfield  {journal}
  {\bibinfo  {journal} {Nature Physics}\ }\textbf {\bibinfo {volume} {10}},\
  \bibinfo {pages} {664--669} (\bibinfo {year} {2014})}\BibitemShut {NoStop}%
\bibitem [{\citenamefont {Schindler}\ \emph {et~al.}(2018)\citenamefont
  {Schindler}, \citenamefont {Wang}, \citenamefont {Vergniory}, \citenamefont
  {Cook}, \citenamefont {Murani}, \citenamefont {Sengupta}, \citenamefont
  {Kasumov}, \citenamefont {Deblock}, \citenamefont {Jeon}, \citenamefont
  {Drozdov}, \citenamefont {Bouchiat}, \citenamefont {Gu{\'e}ron},
  \citenamefont {Yazdani}, \citenamefont {Bernevig},\ and\ \citenamefont
  {Neupert}}]{Schindler2018}%
  \BibitemOpen
  \bibfield  {author} {\bibinfo {author} {\bibfnamefont {F.}~\bibnamefont
  {Schindler}}, \bibinfo {author} {\bibfnamefont {Z.}~\bibnamefont {Wang}},
  \bibinfo {author} {\bibfnamefont {M.~G.}\ \bibnamefont {Vergniory}}, \bibinfo
  {author} {\bibfnamefont {A.~M.}\ \bibnamefont {Cook}}, \bibinfo {author}
  {\bibfnamefont {A.}~\bibnamefont {Murani}}, \bibinfo {author} {\bibfnamefont
  {S.}~\bibnamefont {Sengupta}}, \bibinfo {author} {\bibfnamefont {A.~Y.}\
  \bibnamefont {Kasumov}}, \bibinfo {author} {\bibfnamefont {R.}~\bibnamefont
  {Deblock}}, \bibinfo {author} {\bibfnamefont {S.}~\bibnamefont {Jeon}},
  \bibinfo {author} {\bibfnamefont {I.}~\bibnamefont {Drozdov}}, \bibinfo
  {author} {\bibfnamefont {H.}~\bibnamefont {Bouchiat}}, \bibinfo {author}
  {\bibfnamefont {S.}~\bibnamefont {Gu{\'e}ron}}, \bibinfo {author}
  {\bibfnamefont {A.}~\bibnamefont {Yazdani}}, \bibinfo {author} {\bibfnamefont
  {B.~A.}\ \bibnamefont {Bernevig}}, \ and\ \bibinfo {author} {\bibfnamefont
  {T.}~\bibnamefont {Neupert}},\ }\bibfield  {title} {\enquote {\bibinfo
  {title} {Higher-order topology in bismuth},}\ }\href {\doibase
  10.1038/s41567-018-0224-7} {\bibfield  {journal} {\bibinfo  {journal} {Nature
  Phys.}\ }\textbf {\bibinfo {volume} {14}},\ \bibinfo {pages} {918--924}
  (\bibinfo {year} {2018})}\BibitemShut {NoStop}%
\bibitem [{\citenamefont {Pauly}\ \emph {et~al.}(2015)\citenamefont {Pauly},
  \citenamefont {Rasche}, \citenamefont {Koepernik}, \citenamefont {Liebmann},
  \citenamefont {Pratzer}, \citenamefont {Richter}, \citenamefont {Kellner},
  \citenamefont {Eschbach}, \citenamefont {Kaufmann}, \citenamefont
  {Plucinski}, \citenamefont {Schneider}, \citenamefont {Ruck}, \citenamefont
  {van~den Brink},\ and\ \citenamefont {Morgenstern}}]{Pauly2015}%
  \BibitemOpen
  \bibfield  {author} {\bibinfo {author} {\bibfnamefont {C.}~\bibnamefont
  {Pauly}}, \bibinfo {author} {\bibfnamefont {B.}~\bibnamefont {Rasche}},
  \bibinfo {author} {\bibfnamefont {K.}~\bibnamefont {Koepernik}}, \bibinfo
  {author} {\bibfnamefont {M.}~\bibnamefont {Liebmann}}, \bibinfo {author}
  {\bibfnamefont {M.}~\bibnamefont {Pratzer}}, \bibinfo {author} {\bibfnamefont
  {M.}~\bibnamefont {Richter}}, \bibinfo {author} {\bibfnamefont
  {J.}~\bibnamefont {Kellner}}, \bibinfo {author} {\bibfnamefont
  {M.}~\bibnamefont {Eschbach}}, \bibinfo {author} {\bibfnamefont
  {B.}~\bibnamefont {Kaufmann}}, \bibinfo {author} {\bibfnamefont
  {L.}~\bibnamefont {Plucinski}}, \bibinfo {author} {\bibfnamefont {C.~M.}\
  \bibnamefont {Schneider}}, \bibinfo {author} {\bibfnamefont {M.}~\bibnamefont
  {Ruck}}, \bibinfo {author} {\bibfnamefont {J.}~\bibnamefont {van~den Brink}},
  \ and\ \bibinfo {author} {\bibfnamefont {M.}~\bibnamefont {Morgenstern}},\
  }\bibfield  {title} {\enquote {\bibinfo {title} {Subnanometre-wide electron
  channels protected by topology},}\ }\href {\doibase 10.1038/nphys3264}
  {\bibfield  {journal} {\bibinfo  {journal} {Nature Physics}\ }\textbf
  {\bibinfo {volume} {11}},\ \bibinfo {pages} {338--343} (\bibinfo {year}
  {2015})}\BibitemShut {NoStop}%
\bibitem [{\citenamefont {Sessi}\ \emph {et~al.}(2016)\citenamefont {Sessi},
  \citenamefont {Sante}, \citenamefont {Szczer\-bakow}, \citenamefont {Glott},
  \citenamefont {Wilfert}, \citenamefont {Schmidt}, \citenamefont {Bathon},
  \citenamefont {Dziawa}, \citenamefont {Greiter}, \citenamefont {Neupert},
  \citenamefont {Sangiovanni}, \citenamefont {Story}, \citenamefont {Thomale},\
  and\ \citenamefont {Bode}}]{Sessi2016}%
  \BibitemOpen
  \bibfield  {author} {\bibinfo {author} {\bibfnamefont {P.}~\bibnamefont
  {Sessi}}, \bibinfo {author} {\bibfnamefont {D.~Di}\ \bibnamefont {Sante}},
  \bibinfo {author} {\bibfnamefont {A.}~\bibnamefont {Szczer\-bakow}}, \bibinfo
  {author} {\bibfnamefont {F.}~\bibnamefont {Glott}}, \bibinfo {author}
  {\bibfnamefont {S.}~\bibnamefont {Wilfert}}, \bibinfo {author} {\bibfnamefont
  {H.}~\bibnamefont {Schmidt}}, \bibinfo {author} {\bibfnamefont
  {T.}~\bibnamefont {Bathon}}, \bibinfo {author} {\bibfnamefont
  {P.}~\bibnamefont {Dziawa}}, \bibinfo {author} {\bibfnamefont
  {M.}~\bibnamefont {Greiter}}, \bibinfo {author} {\bibfnamefont
  {T.}~\bibnamefont {Neupert}}, \bibinfo {author} {\bibfnamefont
  {G.}~\bibnamefont {Sangiovanni}}, \bibinfo {author} {\bibfnamefont
  {T.}~\bibnamefont {Story}}, \bibinfo {author} {\bibfnamefont
  {R.}~\bibnamefont {Thomale}}, \ and\ \bibinfo {author} {\bibfnamefont
  {M.}~\bibnamefont {Bode}},\ }\bibfield  {title} {\enquote {\bibinfo {title}
  {Robust spin-polarized midgap states at step edges of topological crystalline
  insulators},}\ }\href {https://science.sciencemag.org/content/354/6317/1269}
  {\bibfield  {journal} {\bibinfo  {journal} {Science}\ }\textbf {\bibinfo
  {volume} {354}},\ \bibinfo {pages} {1269--1273} (\bibinfo {year}
  {2016})}\BibitemShut {NoStop}%
\bibitem [{\citenamefont {Hirahara}\ \emph {et~al.}(2011)\citenamefont
  {Hirahara}, \citenamefont {Bihlmayer}, \citenamefont {Sakamoto},
  \citenamefont {Yamada}, \citenamefont {Miyazaki}, \citenamefont {Kimura},
  \citenamefont {Bl\"ugel},\ and\ \citenamefont
  {Hasegawa}}]{PhysRevLett.107.166801}%
  \BibitemOpen
  \bibfield  {author} {\bibinfo {author} {\bibfnamefont {T.}~\bibnamefont
  {Hirahara}}, \bibinfo {author} {\bibfnamefont {G.}~\bibnamefont {Bihlmayer}},
  \bibinfo {author} {\bibfnamefont {Y.}~\bibnamefont {Sakamoto}}, \bibinfo
  {author} {\bibfnamefont {M.}~\bibnamefont {Yamada}}, \bibinfo {author}
  {\bibfnamefont {H.}~\bibnamefont {Miyazaki}}, \bibinfo {author}
  {\bibfnamefont {S.}~\bibnamefont {Kimura}}, \bibinfo {author} {\bibfnamefont
  {S.}~\bibnamefont {Bl\"ugel}}, \ and\ \bibinfo {author} {\bibfnamefont
  {S.}~\bibnamefont {Hasegawa}},\ }\bibfield  {title} {\enquote {\bibinfo
  {title} {Interfacing {2D} and {3D} topological insulators: {Bi(111)} bilayer
  on {${\mathrm{Bi}}_{2}{\mathrm{Te}}_{3}$}},}\ }\href {\doibase
  10.1103/PhysRevLett.107.166801} {\bibfield  {journal} {\bibinfo  {journal}
  {Phys. Rev. Lett.}\ }\textbf {\bibinfo {volume} {107}},\ \bibinfo {pages}
  {166801} (\bibinfo {year} {2011})}\BibitemShut {NoStop}%
\bibitem [{\citenamefont {Yang}\ \emph {et~al.}(2012)\citenamefont {Yang},
  \citenamefont {Miao}, \citenamefont {Wang}, \citenamefont {Yao},
  \citenamefont {Zhu}, \citenamefont {Song}, \citenamefont {Wang},
  \citenamefont {Xu}, \citenamefont {Fedorov}, \citenamefont {Sun},
  \citenamefont {Zhang}, \citenamefont {Liu}, \citenamefont {Liu},
  \citenamefont {Qian}, \citenamefont {Gao},\ and\ \citenamefont
  {Jia}}]{PhysRevLett.109.016801}%
  \BibitemOpen
  \bibfield  {author} {\bibinfo {author} {\bibfnamefont {F.}~\bibnamefont
  {Yang}}, \bibinfo {author} {\bibfnamefont {L.}~\bibnamefont {Miao}}, \bibinfo
  {author} {\bibfnamefont {Z.~F.}\ \bibnamefont {Wang}}, \bibinfo {author}
  {\bibfnamefont {M.-Y.}\ \bibnamefont {Yao}}, \bibinfo {author} {\bibfnamefont
  {F.}~\bibnamefont {Zhu}}, \bibinfo {author} {\bibfnamefont {Y.~R.}\
  \bibnamefont {Song}}, \bibinfo {author} {\bibfnamefont {M.-X.}\ \bibnamefont
  {Wang}}, \bibinfo {author} {\bibfnamefont {J.-P.}\ \bibnamefont {Xu}},
  \bibinfo {author} {\bibfnamefont {A.~V.}\ \bibnamefont {Fedorov}}, \bibinfo
  {author} {\bibfnamefont {Z.}~\bibnamefont {Sun}}, \bibinfo {author}
  {\bibfnamefont {G.~B.}\ \bibnamefont {Zhang}}, \bibinfo {author}
  {\bibfnamefont {C.}~\bibnamefont {Liu}}, \bibinfo {author} {\bibfnamefont
  {F.}~\bibnamefont {Liu}}, \bibinfo {author} {\bibfnamefont {D.}~\bibnamefont
  {Qian}}, \bibinfo {author} {\bibfnamefont {C.~L.}\ \bibnamefont {Gao}}, \
  and\ \bibinfo {author} {\bibfnamefont {J.-F.}\ \bibnamefont {Jia}},\
  }\bibfield  {title} {\enquote {\bibinfo {title} {Spatial and energy
  distribution of topological edge states in single {Bi(111)} bilayer},}\
  }\href {\doibase 10.1103/PhysRevLett.109.016801} {\bibfield  {journal}
  {\bibinfo  {journal} {Phys. Rev. Lett.}\ }\textbf {\bibinfo {volume} {109}},\
  \bibinfo {pages} {016801} (\bibinfo {year} {2012})}\BibitemShut {NoStop}%
\bibitem [{\citenamefont {Kim}\ \emph {et~al.}(2014)\citenamefont {Kim},
  \citenamefont {Jin}, \citenamefont {Park}, \citenamefont {Kim}, \citenamefont
  {Jhi}, \citenamefont {Kim},\ and\ \citenamefont {Yeom}}]{PhysRevB.89.155436}%
  \BibitemOpen
  \bibfield  {author} {\bibinfo {author} {\bibfnamefont {S.~H.}\ \bibnamefont
  {Kim}}, \bibinfo {author} {\bibfnamefont {K.-H.}\ \bibnamefont {Jin}},
  \bibinfo {author} {\bibfnamefont {J.}~\bibnamefont {Park}}, \bibinfo {author}
  {\bibfnamefont {J.~S.}\ \bibnamefont {Kim}}, \bibinfo {author} {\bibfnamefont
  {S.-H.}\ \bibnamefont {Jhi}}, \bibinfo {author} {\bibfnamefont {T.-H.}\
  \bibnamefont {Kim}}, \ and\ \bibinfo {author} {\bibfnamefont {H.~W.}\
  \bibnamefont {Yeom}},\ }\bibfield  {title} {\enquote {\bibinfo {title} {Edge
  and interfacial states in a two-dimensional topological insulator: {Bi(111)}
  bilayer on {${\mathrm{Bi}}_{2}{\mathrm{Te}}_{2}\mathrm{Se}$}},}\ }\href
  {\doibase 10.1103/PhysRevB.89.155436} {\bibfield  {journal} {\bibinfo
  {journal} {Phys. Rev. B}\ }\textbf {\bibinfo {volume} {89}},\ \bibinfo
  {pages} {155436} (\bibinfo {year} {2014})}\BibitemShut {NoStop}%
\bibitem [{\citenamefont {Hsu}\ \emph {et~al.}(2017)\citenamefont {Hsu},
  \citenamefont {Park},\ and\ \citenamefont {Kim}}]{PhysRevB.96.235433}%
  \BibitemOpen
  \bibfield  {author} {\bibinfo {author} {\bibfnamefont {Y.-T.}\ \bibnamefont
  {Hsu}}, \bibinfo {author} {\bibfnamefont {K.}~\bibnamefont {Park}}, \ and\
  \bibinfo {author} {\bibfnamefont {E.-A.}\ \bibnamefont {Kim}},\ }\bibfield
  {title} {\enquote {\bibinfo {title} {Hybridization-induced interface states
  in a topological-insulator--ferromagnetic-metal heterostructure},}\ }\href
  {\doibase 10.1103/PhysRevB.96.235433} {\bibfield  {journal} {\bibinfo
  {journal} {Phys. Rev. B}\ }\textbf {\bibinfo {volume} {96}},\ \bibinfo
  {pages} {235433} (\bibinfo {year} {2017})}\BibitemShut {NoStop}%
\bibitem [{\citenamefont {Liu}\ \emph {et~al.}(2014)\citenamefont {Liu},
  \citenamefont {Hsieh}, \citenamefont {Wei}, \citenamefont {Duan},
  \citenamefont {Moodera},\ and\ \citenamefont {Fu}}]{LHW2014}%
  \BibitemOpen
  \bibfield  {author} {\bibinfo {author} {\bibfnamefont {J.}~\bibnamefont
  {Liu}}, \bibinfo {author} {\bibfnamefont {T.~H.}\ \bibnamefont {Hsieh}},
  \bibinfo {author} {\bibfnamefont {P.}~\bibnamefont {Wei}}, \bibinfo {author}
  {\bibfnamefont {W.}~\bibnamefont {Duan}}, \bibinfo {author} {\bibfnamefont
  {J.}~\bibnamefont {Moodera}}, \ and\ \bibinfo {author} {\bibfnamefont
  {L.}~\bibnamefont {Fu}},\ }\bibfield  {title} {\enquote {\bibinfo {title}
  {Spin-filtered edge states with an electrically tunable gap in a
  two-dimensional topological crystalline insulator},}\ }\href {\doibase
  10.1038/nmat3828} {\bibfield  {journal} {\bibinfo  {journal} {Nature Mater.}\
  }\textbf {\bibinfo {volume} {13}},\ \bibinfo {pages} {178--183} (\bibinfo
  {year} {2014})}\BibitemShut {NoStop}%
\bibitem [{\citenamefont {Zhang}\ \emph {et~al.}(2010)\citenamefont {Zhang},
  \citenamefont {He}, \citenamefont {Chang}, \citenamefont {Song},
  \citenamefont {Wang}, \citenamefont {Chen}, \citenamefont {Jia},
  \citenamefont {Fang}, \citenamefont {Dai}, \citenamefont {Shan},
  \citenamefont {Shen}, \citenamefont {Niu}, \citenamefont {Qi}, \citenamefont
  {Zhang}, \citenamefont {Ma},\ and\ \citenamefont {Xue}}]{Zhang2010}%
  \BibitemOpen
  \bibfield  {author} {\bibinfo {author} {\bibfnamefont {Y.}~\bibnamefont
  {Zhang}}, \bibinfo {author} {\bibfnamefont {K.}~\bibnamefont {He}}, \bibinfo
  {author} {\bibfnamefont {C.-Z.}\ \bibnamefont {Chang}}, \bibinfo {author}
  {\bibfnamefont {C.-L.}\ \bibnamefont {Song}}, \bibinfo {author}
  {\bibfnamefont {L.-L.}\ \bibnamefont {Wang}}, \bibinfo {author}
  {\bibfnamefont {X.}~\bibnamefont {Chen}}, \bibinfo {author} {\bibfnamefont
  {J.-F.}\ \bibnamefont {Jia}}, \bibinfo {author} {\bibfnamefont
  {Z.}~\bibnamefont {Fang}}, \bibinfo {author} {\bibfnamefont {X.}~\bibnamefont
  {Dai}}, \bibinfo {author} {\bibfnamefont {W.-Y.}\ \bibnamefont {Shan}},
  \bibinfo {author} {\bibfnamefont {S.-Q.}\ \bibnamefont {Shen}}, \bibinfo
  {author} {\bibfnamefont {Q.}~\bibnamefont {Niu}}, \bibinfo {author}
  {\bibfnamefont {X.-L.}\ \bibnamefont {Qi}}, \bibinfo {author} {\bibfnamefont
  {S.-C.}\ \bibnamefont {Zhang}}, \bibinfo {author} {\bibfnamefont {X.-C.}\
  \bibnamefont {Ma}}, \ and\ \bibinfo {author} {\bibfnamefont {Q.-K.}\
  \bibnamefont {Xue}},\ }\bibfield  {title} {\enquote {\bibinfo {title}
  {Crossover of the three-dimensional topological insulator {Bi$_2$Se$_3$} to
  the two-dimensional limit},}\ }\href {\doibase 10.1038/nphys1689} {\bibfield
  {journal} {\bibinfo  {journal} {Nature Phys.}\ }\textbf {\bibinfo {volume}
  {6}},\ \bibinfo {pages} {584--588} (\bibinfo {year} {2010})}\BibitemShut
  {NoStop}%
\bibitem [{\citenamefont {Wojek}\ \emph {et~al.}(2014)\citenamefont {Wojek},
  \citenamefont {Dziawa}, \citenamefont {Kowalski}, \citenamefont
  {Szczerbakow}, \citenamefont {Black-Schaffer}, \citenamefont {Berntsen},
  \citenamefont {Balasubramanian}, \citenamefont {Story},\ and\ \citenamefont
  {Tjernberg}}]{PhysRevB.90.161202}%
  \BibitemOpen
  \bibfield  {author} {\bibinfo {author} {\bibfnamefont {B.~M.}\ \bibnamefont
  {Wojek}}, \bibinfo {author} {\bibfnamefont {P.}~\bibnamefont {Dziawa}},
  \bibinfo {author} {\bibfnamefont {B.~J.}\ \bibnamefont {Kowalski}}, \bibinfo
  {author} {\bibfnamefont {A.}~\bibnamefont {Szczerbakow}}, \bibinfo {author}
  {\bibfnamefont {A.~M.}\ \bibnamefont {Black-Schaffer}}, \bibinfo {author}
  {\bibfnamefont {M.~H.}\ \bibnamefont {Berntsen}}, \bibinfo {author}
  {\bibfnamefont {T.}~\bibnamefont {Balasubramanian}}, \bibinfo {author}
  {\bibfnamefont {T.}~\bibnamefont {Story}}, \ and\ \bibinfo {author}
  {\bibfnamefont {O.}~\bibnamefont {Tjernberg}},\ }\bibfield  {title} {\enquote
  {\bibinfo {title} {Band inversion and the topological phase transition in
  {(Pb,Sn)Se}},}\ }\href {\doibase 10.1103/PhysRevB.90.161202} {\bibfield
  {journal} {\bibinfo  {journal} {Phys. Rev. B}\ }\textbf {\bibinfo {volume}
  {90}},\ \bibinfo {pages} {161202} (\bibinfo {year} {2014})}\BibitemShut
  {NoStop}%
\bibitem [{\citenamefont {Hsieh}\ \emph {et~al.}(2012)\citenamefont {Hsieh},
  \citenamefont {Lin}, \citenamefont {Liu}, \citenamefont {Duan}, \citenamefont
  {Bansil},\ and\ \citenamefont {Fu}}]{Hsieh2012}%
  \BibitemOpen
  \bibfield  {author} {\bibinfo {author} {\bibfnamefont {T.~H.}\ \bibnamefont
  {Hsieh}}, \bibinfo {author} {\bibfnamefont {H.}~\bibnamefont {Lin}}, \bibinfo
  {author} {\bibfnamefont {J.}~\bibnamefont {Liu}}, \bibinfo {author}
  {\bibfnamefont {W.}~\bibnamefont {Duan}}, \bibinfo {author} {\bibfnamefont
  {A.}~\bibnamefont {Bansil}}, \ and\ \bibinfo {author} {\bibfnamefont
  {L.}~\bibnamefont {Fu}},\ }\bibfield  {title} {\enquote {\bibinfo {title}
  {Topological crystalline insulators in the {SnTe} material class},}\ }\href
  {https://www.nature.com/articles/ncomms1969} {\bibfield  {journal} {\bibinfo
  {journal} {Nature Comm.}\ }\textbf {\bibinfo {volume} {3}},\ \bibinfo {pages}
  {982} (\bibinfo {year} {2012})}\BibitemShut {NoStop}%
\bibitem [{\citenamefont {Liu}\ \emph {et~al.}(2013)\citenamefont {Liu},
  \citenamefont {Duan},\ and\ \citenamefont {Fu}}]{Liu2013}%
  \BibitemOpen
  \bibfield  {author} {\bibinfo {author} {\bibfnamefont {J.}~\bibnamefont
  {Liu}}, \bibinfo {author} {\bibfnamefont {W.}~\bibnamefont {Duan}}, \ and\
  \bibinfo {author} {\bibfnamefont {L.}~\bibnamefont {Fu}},\ }\bibfield
  {title} {\enquote {\bibinfo {title} {Two types of surface states in
  topological crystalline insulators},}\ }\href {\doibase
  10.1103/PhysRevB.88.241303} {\bibfield  {journal} {\bibinfo  {journal} {Phys.
  Rev. B}\ }\textbf {\bibinfo {volume} {88}},\ \bibinfo {pages} {241303}
  (\bibinfo {year} {2013})}\BibitemShut {NoStop}%
\bibitem [{\citenamefont {Wang}\ \emph {et~al.}(2013)\citenamefont {Wang},
  \citenamefont {Tsai}, \citenamefont {Lin}, \citenamefont {Xu}, \citenamefont
  {Neupane}, \citenamefont {Hasan},\ and\ \citenamefont {Bansil}}]{Wang2013}%
  \BibitemOpen
  \bibfield  {author} {\bibinfo {author} {\bibfnamefont {Y.~J.}\ \bibnamefont
  {Wang}}, \bibinfo {author} {\bibfnamefont {W.-F.}\ \bibnamefont {Tsai}},
  \bibinfo {author} {\bibfnamefont {H.}~\bibnamefont {Lin}}, \bibinfo {author}
  {\bibfnamefont {S.-Y.}\ \bibnamefont {Xu}}, \bibinfo {author} {\bibfnamefont
  {M.}~\bibnamefont {Neupane}}, \bibinfo {author} {\bibfnamefont {M.~Z.}\
  \bibnamefont {Hasan}}, \ and\ \bibinfo {author} {\bibfnamefont
  {A.}~\bibnamefont {Bansil}},\ }\bibfield  {title} {\enquote {\bibinfo {title}
  {Nontrivial spin texture of the coaxial {Dirac} cones on the surface of
  topological crystalline insulator {SnTe}},}\ }\href {\doibase
  10.1103/PhysRevB.87.235317} {\bibfield  {journal} {\bibinfo  {journal} {Phys.
  Rev. B}\ }\textbf {\bibinfo {volume} {87}},\ \bibinfo {pages} {235317}
  (\bibinfo {year} {2013})}\BibitemShut {NoStop}%
\bibitem [{\citenamefont {Safaei}\ \emph {et~al.}(2013)\citenamefont {Safaei},
  \citenamefont {Kacman},\ and\ \citenamefont {Buczko}}]{Safaei2013}%
  \BibitemOpen
  \bibfield  {author} {\bibinfo {author} {\bibfnamefont {S.}~\bibnamefont
  {Safaei}}, \bibinfo {author} {\bibfnamefont {P.}~\bibnamefont {Kacman}}, \
  and\ \bibinfo {author} {\bibfnamefont {R.}~\bibnamefont {Buczko}},\
  }\bibfield  {title} {\enquote {\bibinfo {title} {Topological crystalline
  insulator {(Pb,Sn)Te}: Surface states and their spin polarization},}\ }\href
  {\doibase 10.1103/PhysRevB.88.045305} {\bibfield  {journal} {\bibinfo
  {journal} {Phys. Rev. B}\ }\textbf {\bibinfo {volume} {88}},\ \bibinfo
  {pages} {045305} (\bibinfo {year} {2013})}\BibitemShut {NoStop}%
\bibitem [{\citenamefont {Okada}\ \emph {et~al.}(2013)\citenamefont {Okada},
  \citenamefont {Serbyn}, \citenamefont {Lin}, \citenamefont {Walkup},
  \citenamefont {Zhou}, \citenamefont {Dhital}, \citenamefont {Neupane},
  \citenamefont {Xu}, \citenamefont {Wang}, \citenamefont {Sankar},
  \citenamefont {Chou}, \citenamefont {Bansil}, \citenamefont {Hasan},
  \citenamefont {Wilson}, \citenamefont {Fu},\ and\ \citenamefont
  {Madhavan}}]{Okada2013}%
  \BibitemOpen
  \bibfield  {author} {\bibinfo {author} {\bibfnamefont {Y.}~\bibnamefont
  {Okada}}, \bibinfo {author} {\bibfnamefont {M.}~\bibnamefont {Serbyn}},
  \bibinfo {author} {\bibfnamefont {H.}~\bibnamefont {Lin}}, \bibinfo {author}
  {\bibfnamefont {D.}~\bibnamefont {Walkup}}, \bibinfo {author} {\bibfnamefont
  {W.}~\bibnamefont {Zhou}}, \bibinfo {author} {\bibfnamefont {C.}~\bibnamefont
  {Dhital}}, \bibinfo {author} {\bibfnamefont {M.}~\bibnamefont {Neupane}},
  \bibinfo {author} {\bibfnamefont {S.}~\bibnamefont {Xu}}, \bibinfo {author}
  {\bibfnamefont {Y.~J.}\ \bibnamefont {Wang}}, \bibinfo {author}
  {\bibfnamefont {R.}~\bibnamefont {Sankar}}, \bibinfo {author} {\bibfnamefont
  {F.}~\bibnamefont {Chou}}, \bibinfo {author} {\bibfnamefont {A.}~\bibnamefont
  {Bansil}}, \bibinfo {author} {\bibfnamefont {M.~Z.}\ \bibnamefont {Hasan}},
  \bibinfo {author} {\bibfnamefont {S.~D.}\ \bibnamefont {Wilson}}, \bibinfo
  {author} {\bibfnamefont {L.}~\bibnamefont {Fu}}, \ and\ \bibinfo {author}
  {\bibfnamefont {V.}~\bibnamefont {Madhavan}},\ }\bibfield  {title} {\enquote
  {\bibinfo {title} {Observation of {Dirac} node formation and mass acquisition
  in a topological crystalline insulator},}\ }\href {\doibase
  10.1126/science.1239451} {\bibfield  {journal} {\bibinfo  {journal}
  {Science}\ }\textbf {\bibinfo {volume} {341}},\ \bibinfo {pages} {1496--1499}
  (\bibinfo {year} {2013})}\BibitemShut {NoStop}%
\bibitem [{Sup()}]{SupplMat}%
  \BibitemOpen
  \href@noop {} {}\bibinfo {note} {See supplemental material for detailed
  information regarding the electronic structure of odd- and even-atomic step
  edges, details regarding crystal synthesis and characterization, further data
  obtained on wedge-shaped islands and depressions, the asymmetric energy
  splitting observed for two parallel single-atomic steps, our model of two
  interacting 1D edge modes, single-atomic steps crossing under angles of about
  $60^{\circ}$ and $90^{\circ}$, and fluctuations of the Dirac
  energy.}\BibitemShut {Stop}%
\bibitem [{\citenamefont {Iaia}\ \emph {et~al.}(2019)\citenamefont {Iaia},
  \citenamefont {Wang}, \citenamefont {Maximenko}, \citenamefont {Walkup},
  \citenamefont {Sankar}, \citenamefont {Chou}, \citenamefont {Lu},\ and\
  \citenamefont {Madhavan}}]{Madhavan2019}%
  \BibitemOpen
  \bibfield  {author} {\bibinfo {author} {\bibfnamefont {D.}~\bibnamefont
  {Iaia}}, \bibinfo {author} {\bibfnamefont {C.-Y.}\ \bibnamefont {Wang}},
  \bibinfo {author} {\bibfnamefont {Y.}~\bibnamefont {Maximenko}}, \bibinfo
  {author} {\bibfnamefont {D.}~\bibnamefont {Walkup}}, \bibinfo {author}
  {\bibfnamefont {R.}~\bibnamefont {Sankar}}, \bibinfo {author} {\bibfnamefont
  {F.}~\bibnamefont {Chou}}, \bibinfo {author} {\bibfnamefont {Y.-M.}\
  \bibnamefont {Lu}}, \ and\ \bibinfo {author} {\bibfnamefont {V.}~\bibnamefont
  {Madhavan}},\ }\bibfield  {title} {\enquote {\bibinfo {title} {Topological
  nature of step-edge states on the surface of the topological crystalline
  insulator {${\mathrm{Pb}}_{0.7}{\mathrm{Sn}}_{0.3}\mathrm{Se}$}},}\ }\href
  {\doibase 10.1103/PhysRevB.99.155116} {\bibfield  {journal} {\bibinfo
  {journal} {Phys. Rev. B}\ }\textbf {\bibinfo {volume} {99}},\ \bibinfo
  {pages} {155116} (\bibinfo {year} {2019})}\BibitemShut {NoStop}%
\bibitem [{\citenamefont {Rechci\ifmmode~\acute{n}\else \'{n}\fi{}ski}\ and\
  \citenamefont {Buczko}(2018)}]{Rechcinski2018}%
  \BibitemOpen
  \bibfield  {author} {\bibinfo {author} {\bibfnamefont {R.}~\bibnamefont
  {Rechci\ifmmode~\acute{n}\else \'{n}\fi{}ski}}\ and\ \bibinfo {author}
  {\bibfnamefont {R.}~\bibnamefont {Buczko}},\ }\bibfield  {title} {\enquote
  {\bibinfo {title} {Topological states on uneven {(Pb,Sn)Se} (001)
  surfaces},}\ }\href {\doibase 10.1103/PhysRevB.98.245302} {\bibfield
  {journal} {\bibinfo  {journal} {Phys. Rev. B}\ }\textbf {\bibinfo {volume}
  {98}},\ \bibinfo {pages} {245302} (\bibinfo {year} {2018})}\BibitemShut
  {NoStop}%
\bibitem [{\citenamefont {Beidenkopf}\ \emph {et~al.}(2011)\citenamefont
  {Beidenkopf}, \citenamefont {Roushan}, \citenamefont {Seo}, \citenamefont
  {Gorman}, \citenamefont {Drozdov}, \citenamefont {Hor}, \citenamefont
  {Cava},\ and\ \citenamefont {Yazdani}}]{Beidenkopf2011}%
  \BibitemOpen
  \bibfield  {author} {\bibinfo {author} {\bibfnamefont {H.}~\bibnamefont
  {Beidenkopf}}, \bibinfo {author} {\bibfnamefont {P.}~\bibnamefont {Roushan}},
  \bibinfo {author} {\bibfnamefont {J.}~\bibnamefont {Seo}}, \bibinfo {author}
  {\bibfnamefont {L.}~\bibnamefont {Gorman}}, \bibinfo {author} {\bibfnamefont
  {I.}~\bibnamefont {Drozdov}}, \bibinfo {author} {\bibfnamefont {Y.~S.}\
  \bibnamefont {Hor}}, \bibinfo {author} {\bibfnamefont {R.~J.}\ \bibnamefont
  {Cava}}, \ and\ \bibinfo {author} {\bibfnamefont {A.}~\bibnamefont
  {Yazdani}},\ }\bibfield  {title} {\enquote {\bibinfo {title} {Spatial
  fluctuations of helical {Dirac} fermions on the surface of topological
  insulators},}\ }\href {\doibase 10.1038/nphys2108} {\bibfield  {journal}
  {\bibinfo  {journal} {Nature Phys.}\ }\textbf {\bibinfo {volume} {7}},\
  \bibinfo {pages} {939--943} (\bibinfo {year} {2011})}\BibitemShut {NoStop}%
\bibitem [{\citenamefont {Storz}\ \emph {et~al.}(2016)\citenamefont {Storz},
  \citenamefont {Cortijo}, \citenamefont {Wilfert}, \citenamefont {Kokh},
  \citenamefont {Te\-reshchenko}, \citenamefont {Vozmediano}, \citenamefont
  {Bode}, \citenamefont {Guinea},\ and\ \citenamefont {Sessi}}]{Storz2016}%
  \BibitemOpen
  \bibfield  {author} {\bibinfo {author} {\bibfnamefont {O.}~\bibnamefont
  {Storz}}, \bibinfo {author} {\bibfnamefont {A.}~\bibnamefont {Cortijo}},
  \bibinfo {author} {\bibfnamefont {S.}~\bibnamefont {Wilfert}}, \bibinfo
  {author} {\bibfnamefont {K.~A.}\ \bibnamefont {Kokh}}, \bibinfo {author}
  {\bibfnamefont {O.~E.}\ \bibnamefont {Te\-reshchenko}}, \bibinfo {author}
  {\bibfnamefont {M.~A.~H.}\ \bibnamefont {Vozmediano}}, \bibinfo {author}
  {\bibfnamefont {M.}~\bibnamefont {Bode}}, \bibinfo {author} {\bibfnamefont
  {F.}~\bibnamefont {Guinea}}, \ and\ \bibinfo {author} {\bibfnamefont
  {P.}~\bibnamefont {Sessi}},\ }\bibfield  {title} {\enquote {\bibinfo {title}
  {Mapping the effect of defect-induced strain disorder on the {Dirac} states
  of topological insulators},}\ }\href {\doibase 10.1103/PhysRevB.94.121301}
  {\bibfield  {journal} {\bibinfo  {journal} {Phys. Rev. B}\ }\textbf {\bibinfo
  {volume} {94}},\ \bibinfo {pages} {121301} (\bibinfo {year}
  {2016})}\BibitemShut {NoStop}%
\bibitem [{\citenamefont {Nilsson}\ \emph {et~al.}(2018)\citenamefont
  {Nilsson}, \citenamefont {Bostr\"om}, \citenamefont {Lehmann}, \citenamefont
  {Dick}, \citenamefont {Leijnse},\ and\ \citenamefont
  {Thelander}}]{Nilson2018}%
  \BibitemOpen
  \bibfield  {author} {\bibinfo {author} {\bibfnamefont {M.}~\bibnamefont
  {Nilsson}}, \bibinfo {author} {\bibfnamefont {F.~V.}\ \bibnamefont
  {Bostr\"om}}, \bibinfo {author} {\bibfnamefont {S.}~\bibnamefont {Lehmann}},
  \bibinfo {author} {\bibfnamefont {K.~A.}\ \bibnamefont {Dick}}, \bibinfo
  {author} {\bibfnamefont {M.}~\bibnamefont {Leijnse}}, \ and\ \bibinfo
  {author} {\bibfnamefont {C.}~\bibnamefont {Thelander}},\ }\bibfield  {title}
  {\enquote {\bibinfo {title} {Tuning the two-electron hybridization and spin
  states in parallel-coupled {InAs} quantum dots},}\ }\href {\doibase
  10.1103/PhysRevLett.121.156802} {\bibfield  {journal} {\bibinfo  {journal}
  {Phys. Rev. Lett.}\ }\textbf {\bibinfo {volume} {121}},\ \bibinfo {pages}
  {156802} (\bibinfo {year} {2018})}\BibitemShut {NoStop}%
\bibitem [{\citenamefont {Coy~Diaz}\ \emph {et~al.}(2015)\citenamefont
  {Coy~Diaz}, \citenamefont {Avila}, \citenamefont {Chen}, \citenamefont
  {Addou}, \citenamefont {Asensio},\ and\ \citenamefont {Batzill}}]{Diaz2015}%
  \BibitemOpen
  \bibfield  {author} {\bibinfo {author} {\bibfnamefont {H.}~\bibnamefont
  {Coy~Diaz}}, \bibinfo {author} {\bibfnamefont {J.}~\bibnamefont {Avila}},
  \bibinfo {author} {\bibfnamefont {C.}~\bibnamefont {Chen}}, \bibinfo {author}
  {\bibfnamefont {R.}~\bibnamefont {Addou}}, \bibinfo {author} {\bibfnamefont
  {M.~C.}\ \bibnamefont {Asensio}}, \ and\ \bibinfo {author} {\bibfnamefont
  {M.}~\bibnamefont {Batzill}},\ }\bibfield  {title} {\enquote {\bibinfo
  {title} {Direct observation of interlayer hybridization and {D}irac
  relativistic carriers in graphene/{MoS$_2$} van der {W}aals
  heterostructures},}\ }\href {\doibase 10.1021/nl504167y} {\bibfield
  {journal} {\bibinfo  {journal} {Nano Letters}\ }\textbf {\bibinfo {volume}
  {15}},\ \bibinfo {pages} {1135--1140} (\bibinfo {year} {2015})}\BibitemShut
  {NoStop}%
\bibitem [{Note1()}]{Note1}%
  \BibitemOpen
  \bibinfo {note} {We would like to note that we cannot exclude that similar or
  even better results can be obtained by more elaborate models, where the
  electrostatic interaction becomes less relevant and is instead replaced by a
  behavior governed by spatially varying topological indices}\BibitemShut
  {NoStop}%
\bibitem [{\citenamefont {Zeljkovic}\ \emph {et~al.}(2014)\citenamefont
  {Zeljkovic}, \citenamefont {Okada}, \citenamefont {Huang}, \citenamefont
  {Sankar}, \citenamefont {Walkup}, \citenamefont {Zhou}, \citenamefont
  {Serbyn}, \citenamefont {Chou}, \citenamefont {Tsai}, \citenamefont {Lin},
  \citenamefont {Bansil}, \citenamefont {Fu}, \citenamefont {Hasan},\ and\
  \citenamefont {Madhavan}}]{Zeljkovic2014}%
  \BibitemOpen
  \bibfield  {author} {\bibinfo {author} {\bibfnamefont {I.}~\bibnamefont
  {Zeljkovic}}, \bibinfo {author} {\bibfnamefont {Y.}~\bibnamefont {Okada}},
  \bibinfo {author} {\bibfnamefont {C.-Y.}\ \bibnamefont {Huang}}, \bibinfo
  {author} {\bibfnamefont {R.}~\bibnamefont {Sankar}}, \bibinfo {author}
  {\bibfnamefont {D.}~\bibnamefont {Walkup}}, \bibinfo {author} {\bibfnamefont
  {W.}~\bibnamefont {Zhou}}, \bibinfo {author} {\bibfnamefont {M.}~\bibnamefont
  {Serbyn}}, \bibinfo {author} {\bibfnamefont {F.}~\bibnamefont {Chou}},
  \bibinfo {author} {\bibfnamefont {W.-F.}\ \bibnamefont {Tsai}}, \bibinfo
  {author} {\bibfnamefont {H.}~\bibnamefont {Lin}}, \bibinfo {author}
  {\bibfnamefont {A.}~\bibnamefont {Bansil}}, \bibinfo {author} {\bibfnamefont
  {L.}~\bibnamefont {Fu}}, \bibinfo {author} {\bibfnamefont {M.~Z.}\
  \bibnamefont {Hasan}}, \ and\ \bibinfo {author} {\bibfnamefont
  {V.}~\bibnamefont {Madhavan}},\ }\bibfield  {title} {\enquote {\bibinfo
  {title} {Mapping the unconventional orbital texture in topological
  crystalline insulators},}\ }\href {\doibase 10.1038/nphys3012} {\bibfield
  {journal} {\bibinfo  {journal} {Nature Phys.}\ }\textbf {\bibinfo {volume}
  {10}},\ \bibinfo {pages} {572--577} (\bibinfo {year} {2014})}\BibitemShut
  {NoStop}%
\bibitem [{\citenamefont {Wojek}\ \emph {et~al.}(2013)\citenamefont {Wojek},
  \citenamefont {Buczko}, \citenamefont {Safaei}, \citenamefont {Dziawa},
  \citenamefont {Kowalski}, \citenamefont {Berntsen}, \citenamefont
  {Balasubramanian}, \citenamefont {Leandersson}, \citenamefont {Szczerbakow},
  \citenamefont {Kacman}, \citenamefont {Story},\ and\ \citenamefont
  {Tjernberg}}]{PhysRevB.87.115106}%
  \BibitemOpen
  \bibfield  {author} {\bibinfo {author} {\bibfnamefont {B.~M.}\ \bibnamefont
  {Wojek}}, \bibinfo {author} {\bibfnamefont {R.}~\bibnamefont {Buczko}},
  \bibinfo {author} {\bibfnamefont {S.}~\bibnamefont {Safaei}}, \bibinfo
  {author} {\bibfnamefont {P.}~\bibnamefont {Dziawa}}, \bibinfo {author}
  {\bibfnamefont {B.~J.}\ \bibnamefont {Kowalski}}, \bibinfo {author}
  {\bibfnamefont {M.~H.}\ \bibnamefont {Berntsen}}, \bibinfo {author}
  {\bibfnamefont {T.}~\bibnamefont {Balasubramanian}}, \bibinfo {author}
  {\bibfnamefont {M.}~\bibnamefont {Leandersson}}, \bibinfo {author}
  {\bibfnamefont {A.}~\bibnamefont {Szczerbakow}}, \bibinfo {author}
  {\bibfnamefont {P.}~\bibnamefont {Kacman}}, \bibinfo {author} {\bibfnamefont
  {T.}~\bibnamefont {Story}}, \ and\ \bibinfo {author} {\bibfnamefont
  {O.}~\bibnamefont {Tjernberg}},\ }\bibfield  {title} {\enquote {\bibinfo
  {title} {Spin-polarized (001) surface states of the topological crystalline
  insulator {Pb${}_{0.73}$Sn${}_{0.27}$Se}},}\ }\href {\doibase
  10.1103/PhysRevB.87.115106} {\bibfield  {journal} {\bibinfo  {journal} {Phys.
  Rev. B}\ }\textbf {\bibinfo {volume} {87}},\ \bibinfo {pages} {115106}
  (\bibinfo {year} {2013})}\BibitemShut {NoStop}%
\end{thebibliography}

%

\end{document}